\documentclass[aps,prd,preprint,superscriptaddress,showpacs,longbibliography,footnoteadded]{revtex4-1}
\usepackage[T1]{fontenc}
\usepackage{float}
\usepackage{subfigure}
\usepackage{amssymb}
\usepackage{graphicx}
\usepackage{amsmath}
\usepackage{slashed}
\usepackage{tensor}
\usepackage{hyperref}
\usepackage{epstopdf}
\usepackage{extarrows}
\usepackage{appendix}
\usepackage{color}
\usepackage{bbm}
\usepackage{boldline}
\usepackage{caption}%caption 宏包，加强对caption的控制，应该主要用于给table头顶加一个caption吧，一旦启用此package，图表的caption自动中间对齐
\usepackage[section]{placeins}%可以禁止图片或者表格浮动超过\section的范围

\usepackage[utf8]{inputenc}
\hypersetup{
    colorlinks=true,
    linkcolor=red,
    citecolor=blue,
}

\usepackage{orcidlink}

\begin{document}

\title{Scalar perturbations and strong cosmic censorship for an asymptotically de Sitter regular black hole}
\author{Hang Liu\orcidlink{0000-0003-1582-3977}}
\email{hangliu@sjtu.edu.cn}
\affiliation{College of Physics and Materials Science, Tianjin Normal University, Tianjin 300387, China}

\author{Hong Guo\orcidlink{0000-0002-0999-684X}}
\email{hong_guo@pku.edu.cn}
\affiliation{Kavli Institute for Astronomy and Astrophysics, Peking University, Beijing 100871, China}

%%%%%%%%%%%%%%%%%%%%%%%%%%%%%%%%%%%%%%%%%%%%%%%%%%%%%%%%%%%%%%%%%%%%%%%%%%%%
\begin{abstract}

We investigate the quasinormal modes (QNMs) of scalar perturbations and strong cosmic censorship (SCC) in regular Ayón-Beato-García-Bronnikov-de Sitter (ABGB-dS) black hole spacetime. The main motivation of this work is to clarify whether the regularization of black hole core, which removes the central singularity, can also change the fate of SCC connected to the Cauchy horizon. We first study the dependence of the scalar QNM frequency (QNF) on the scalar mass and black hole parameters. For angular number  $l=0$, a small  scalar mass can give rise to a slowly decaying mode which is purely imaginary, while it will become complex for large enough  scalar mass and we call it dS-like mode. We also find that the black hole charge and cosmological constant affect the QNM spectrum in qualitatively opposite ways.
We then examine SCC under massless and massive scalar perturbations. For massless perturbations, the QNMs spectrum can be organized into photon sphere (PS), de Sitter (dS) and near-extremal (NE) families. Compared with the SCC  in the Reissner-Nordström-de Sitter (RN-dS) case, SCC is generally more easily violated in the ABGB-dS spacetime, although this tendency is reversed for sufficiently large cosmological constant. For massive scalar perturbations, we find that the scalar mass does not restore SCC. Instead, it can enhance the violation by driving the system into the regime $\beta>1$. At last, we discuss what we may expect from the violation of SCC in a regular black hole spacetime.
\end{abstract}

\date{\today}

\maketitle

%%%%%%%%%%%%%%%%%%%%%%%%%%%%%%%%%%%%%%%%%%%%%%%%%%%%%%%%%%%%%%%%%%%%%%%%%%%%
\section{Introduction}\label{sec1}

The predictability is one of the most important features of physics at classical level, including Einstein's general relativity (GR) whose deterministic character is clearly formulated through its initial value problem. Once the initial data are appropriately given, a maximal globally hyperbolic development (MGHD) of spacetime is determined by Einstein's equations. However, the possible existence of Cauchy horizon  makes GR lose its global predictability, since the Cauchy horizon is the boundary of the  spacetime beyond which the spacetime evolution  cannot be  uniquely determined by the initial data posed on a hypersurface. In other words, a failure of determinism at the classical level would  arise if the MGHD is extendible as a solution to  Einstein's equations. In light of this, SCC is  proposed  to save the predictability of GR. The strongest version of SCC is the $C^0$ formulation: The MGHD is inextendible as a $C^0$ Lorentzian manifold \cite{VandeMoortel:2025ngd}. However, it has been proven that the metric of charged black hole spacetime can be continuously extended across the Cauchy horizon \cite{Dafermos:2003wr,Dafermos:2017dbw}, thereby falsifying the $C^0$ version of SCC. Nevertheless, the extended spacetime may be unphysical if the metric is only continuous across the Cauchy horizon but lacks any other regularity requirements, since Einstein's equations may not be satisfied by this extended metric, even in a weak sense. 

On the other hand, the extended object may be able to  cross the Cauchy horizon with divergent curvature without being destroyed, which is contrary to the intuition but it indeed can happen if the curvature is weak enough \cite{Ori:1991zz}. Although the divergent curvature indicates that the metric is not $C^2$, it may still solve the Einstein's equations which are second order as a weak solution, which poses a potential threat to the determinism of GR. Therefore, to protect the predictability of GR, we have to require the metric to be inextendible across the Cauchy horizon as a weak solution to the Einstein's equations. This observation brings us to the Christodoulou’s modern formulation of SCC (or $H^1$ formulation, weaker strength than $C^0$ formulation) which asserts that the metric is possible to be continuously extended across the  Cauchy horizon, but this should not be done with locally square integrable Christoﬀel symbols \cite{Christodoulou:2008nj}. We should point out that the original SCC proposed by Penrose concerns the problem of singularities by conjecturing that the naked or locally naked singularities should be absent in generic gravitational collapse \cite{Penrose:1980ge,1974IAUS...64...82P}, which differs from the modern formulation of SCC which concerns the extendibility of metric across Cauchy horizon. One can refer to \cite{VandeMoortel:2025ngd} for a recent  review of this topic.

The nonlinear analysis in \cite{Hintz:2016gwb,Hintz:2016jak,Costa:2017tjc,Dafermos:2017dbw} suggests that  the metric and scalar field have similar extendibility properties, which allows us to use the scalar field as a proxy of extendibility of the spacetime metric. Hence, instead of studying SCC directly, we can use linear scalar perturbations as a toy model by which we have an analogue of the Christodoulou's formulation of SCC stating that the scalar fields and their gradient  cannot be locally square integrable at the Cauchy horizon, which means that the scalar fields are required not to belong to the Sobolev space $H^1_{\rm loc}$ (higher regularity may be required for some scenarios, see examples in \cite{Destounis:2019omd} which also explicitly shows that the scalar field and spacetime metric share similar regularity requirements). Recalling the form of standard energy-momentum tensor of minimally coupled scalar fields, the physical consequence of this requirement is that the energy of scalar fields will diverge at Cauchy horizon. This version of SCC has been found to hold in RN and Kerr black hole spacetime with cosmological constant $\Lambda=0$ \cite{Luk:2015qja,Dafermos:2015bzz}. However, for a spacetime with a $\Lambda>0$, the validity of SCC may be in jeopardy. To understand this, we need to know that the SCC is related to the instability of Cauchy horizon. When late time scalar perturbations propagate along the Cauchy horizon, they will suffer a blueshift effect which may lead to a mass-inflation thereby leading to instability of Cauchy horizon and providing a basic condition for the SCC to be respected. This is indeed the case for spacetime with $\Lambda=0$ where the decay of perturbations follows a power law which does not have  a fast enough decay rate to compete with the blueshift amplification effect on Cauchy horizon. 

However, it is known that the late time perturbations $\phi$ decay exponentially by $|\phi-\phi_0|\leq Ce^{-\alpha t}$ ($\phi_0\in \mathbb{C}$ is some constant, and $\alpha$ stands for spectral gap and can be determined by the dominant QNMs) for $\Lambda>0$ since the perturbations can fall into cosmological horizon. This sufficiently fast decay rate of perturbations may lead to a weaker instability of Cauchy horizon, therefore hinting a potential failure of Christodoulou's formulation of SCC. The competition between the amplification and decay effects depends on $\beta\equiv\alpha/\kappa_-$, where $\kappa_-$ is the surface gravity of Cauchy horizon. Then we can claim the violation of the Christodoulou's formulation of SCC  if we find $\beta>1/2$. In fact, a  recent remarkable progress from Cardoso \cite{Cardoso:2017soq} argued  that the SCC is indeed violated under massless scalar perturbations for a near-extremal RN-dS black hole by providing the evidence of $\beta>1/2$.
Undoubtedly,  Cardoso's work \cite{Cardoso:2017soq} has revived the intense interests of studying SCC and considerable efforts have been put into this topic \cite{Cardoso:2018nvb,Dias:2018ufh,Dias:2018ynt,Dias:2018etb,Hod:2018dpx,Destounis:2018qnb,Rahman:2018oso,Mo:2018nnu,Ge:2018vjq,Dias:2019ery,Liu:2019lon,Guo:2019tjy,Liu:2019rbq,Destounis:2019omd,Mishra:2020jlw,Gan:2019jac,Emparan:2020rnp,Singha:2022bvr,Konoplya:2022kld,Zhang:2023yco,Courty:2023rxk,Davey:2024xvd,Lin:2024beb,Chrysostomou:2025qud,Tu:2025zeb,Li:2026zsg}. 

The modern formulation of SCC, framed as an initial value problem, does not concern itself with anything inside the Cauchy horizon, including whether a central singularity exists. Most of the works mentioned above have focused on black holes harboring curvature singularities, while the SCC in regular black hole spacetimes has received far less attention. A likely reason for this situation is the long-standing conceptual association between cosmic censorship and the presence of a central singularity. Historically, discussions of cosmic censorship were closely tied to the singularities formed in gravitational collapse, which may leave the impression that regular black holes fall outside the scope of SCC. But this impression does not survive a careful reading of the modern formulation. It asks whether the MGHD, the maximal region of unique evolution dictated by the initial data, can be extended across the Cauchy horizon as a weak solution of the field equations, and this question is posed entirely from the domain-of-dependence side. In this respect, regular black holes are squarely within SCC's domain of applicability. It is instead the weak cosmic censorship that loses its relevance in singularity-free spacetimes, since what it concerns is the visibility of singularities to distant observers.

In fact, general analyses of nonsingular black hole geometries have shown that, under rather mild assumptions, replacing the central singularity with a regular core generically leads to the appearance of an inner horizon \cite{Frolov:2016pav,Carballo-Rubio:2019fnb}. The stability of this inner horizon in regular black holes is currently under active debate. On one side, it has been argued that the mass inflation mechanism originally discovered in the Reissner-Nordström interior \cite{Poisson:1990eh} operates on any inner horizon with nonzero surface gravity, thereby destabilizing the core of regular black holes \cite{Carballo-Rubio:2018pmi,Carballo-Rubio:2021bpr}. On the other side, it has been claimed that once the late-time attractor of the perturbed geometry is properly taken into account, regular black holes possess stable cores \cite{Bonanno:2020fgp,Bonanno:2022jjp}. Meanwhile, regular geometries that circumvent mass inflation altogether have been constructed, such as models with vanishing inner-horizon surface gravity \cite{Carballo-Rubio:2022kad,Franzin:2022wai}. So far, no consensus has been reached on the fate of the inner horizon of regular black holes. The SCC analysis in asymptotically dS spacetimes enters this debate from a complementary direction: the parameter $\beta = \alpha/\kappa_-$ precisely captures the competition between blueshift amplification at the Cauchy horizon and the fastest possible decay rate set by the spectral gap, thereby providing an independent and sharp diagnostic of the strength of the Cauchy horizon instability for regular black holes.

On the other hand, SCC in regular black holes connects directly to broader questions in quantum gravity. Regular black holes are often introduced as low-energy effective descriptions of quantum gravity, and their Cauchy horizon physics probes whether quantum effects restore or further undermine classical determinism. Recent studies have investigated mass inflation and SCC in regular black holes inspired by loop quantum gravity \cite{Cao:2023aco,Liu:2026ltw} and covariant quantum gravity \cite{Lin:2024beb}, finding that quantum corrections do not trivially remove the Cauchy horizon instability. If sufficiently regular extensions across the Cauchy horizon persist after gravitational backreaction is taken into consideration, the continuation of the spacetime may connect to the regular interior rather than terminate at a curvature singularity. In this sense, a violation of SCC in a regular black hole may have implications for the accessibility of the regular core.  The very result that threatens determinism would then also point to a physically accessible route into the region usually expected to encode the quantum gravity effects responsible for resolving the singularity \cite{Ashtekar:2005qt,Bojowald:2018xxu,Ashtekar:2018lag}. Hence, a violation of SCC in regular black hole spacetime is therefore not merely a pathology; it may also carry information about the accessibility of the regular core, provided that the stability and extendibility  of  the Cauchy horizon still persist when nonlinear effects are considered.

Furthermore, whether or not SCC is violated in regular black holes may carry observable astrophysical imprints. Cao et al.\ \cite{Cao:2023par} studied the appearance of a regular black hole with a stable inner horizon, which violates SCC, and showed that in the maximally extended spacetime the photons entering the horizons of the black hole in the preceding companion universe can traverse the stable Cauchy horizon and emerge from the white hole region into our universe, imprinting a distinctive multi-ring structure on the black hole image. Moreover, since regular black holes are not shielded by the weak cosmic censorship conjecture, their event horizons can in principle be destroyed \cite{Li:2013sea,Yang:2022yvq}, leaving horizonless configurations whose regular cores become accessible to light rays, with observational signatures that are beginning to be explored \cite{Fauzi:2025ldu}. Should the SCC fate differ between singular and regular black holes, or vary among different regular black hole models, this would offer an observational pathway to probe black hole interior structure through the gravitational wave ringdown, the QNM spectrum, or the electromagnetic images of black holes.

The above considerations motivate a systematic study of SCC in regular black hole spacetime. The regular ABGB-dS black hole \cite{Matyjasek:2008kn}, which is charged and supported by nonlinear electrodynamics, offers us a simple framework in which we can address the questions relevant to SCC. For appropriate parameter region, it can possess three horizons, namely cosmological horizon, event horizon and inner Cauchy horizon, just like other charged black holes in asymptotically dS spacetime. It has been found that the asymptotic geometry of ABGB-dS solution resembles the RN-dS spacetime, while its near-horizon dynamics is noticeably modified by  the nonlinear electrodynamics. It is therefore important to determine whether the properties of SCC revealed for RN-dS case persist in this regular black hole spacetime and whether some qualitatively new characteristics emerge. It should be noted that the SCC we are about to study refers to the Christodoulou's formulation of SCC. We will start from the analysis of the properties of QNMs spectra, not only due to the spectral gap used in the confirmation of violation of SCC is determined by the lowest lying QNF, but the QNMs also reflect the dynamical properties of the black hole in consideration.  

The rest of this work is organized as follows. In Sec.~\ref{sec2}, we introduce the ABGB-dS black hole solution, scalar perturbations and numerical methods. In Sec.~\ref{sec3}, we discuss the numerical results of QNF, including massless and massive scalar QNMs. In Sec.~\ref{sec4}, we investigate the SCC and make comparison with RN-dS case. As the last section, Sec.~\ref{sec5} is devoted to conclusions and discussions.

\section{Basic Formulas and Numerical Methods} \label{sec2}

In this section, we would like to give a brief introduction to the regular ABGB-dS black hole solution and the numerical methods adopted in our calculation of QNF of scalar field perturbations. 

\subsection{ABGB-dS black hole solution and scalar perturbations}

\begin{figure}[thbp]
\centering
\includegraphics[height=2.8in,width=4.1in]{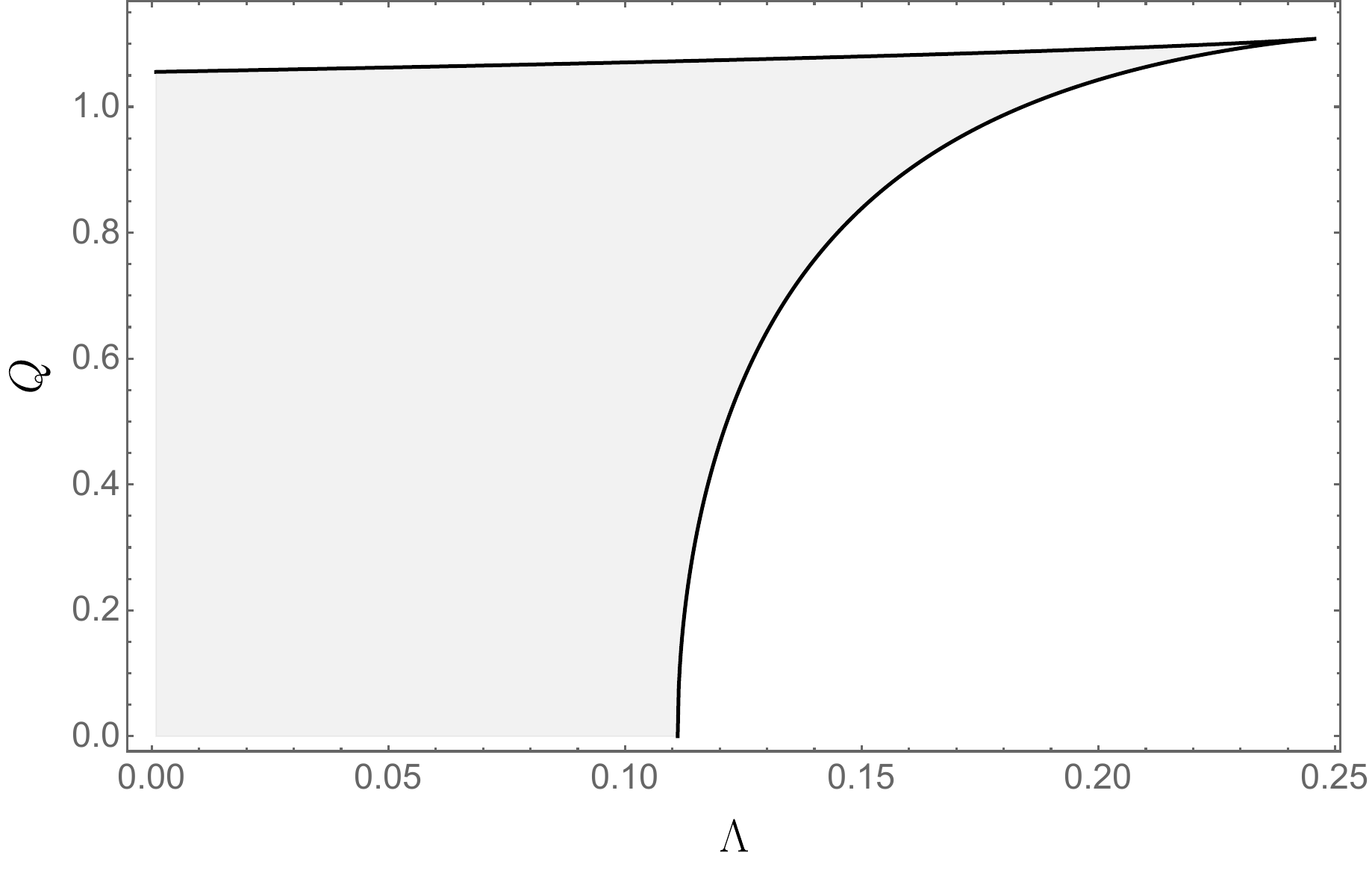}
\caption{The parameter space of ABGB-dS black hole. The interior of the parameter space (shaded area) represents black hole parameters which guarantee three distinct horizons, while at most two horizons are allowed on the boundary. We have $r_-=r_+$ on the upper  black boundary  curve, and $r_+=r_c$ on the lower black boundary curve.\label{fig0}}
\end{figure}

Considering a spacetime with a  positive cosmological constant $\Lambda$ and the gravity is coupled to nonlinear electrodynamics, the corresponding action is given by \cite{Matyjasek:2008kn}

\begin{equation}
	S=\frac{1}{16\pi}\int d^4x \sqrt{-g}\left(R-2\Lambda-\mathcal{L}(F)\right),
\end{equation}
where $\mathcal{L}(F)$ is the Lagrangian of nonlinear electrodynamics, which has been chosen to be \cite{Matyjasek:2008kn}

\begin{equation}
\mathcal{L}(F)=F\left[1-\tanh^2\left(s\sqrt[4]{\frac{Q^2F}{2}}\right)\right], \quad F=F_{ab}F^{ab}=\frac{2Q^2}{r^4}, \quad s=\frac{|Q|}{2b},
\end{equation}
where $Q$ is interpreted as magnetic charge. The free parameter $b$ is tuned to some value by which the center of the black hole will be regular. A Static spherically symmetric solution for this system is given by ABGB-dS black hole metric \cite{Matyjasek:2008kn}

\begin{equation}
	ds^2=-f(r)dt^2+\frac{dr^2}{f(r)}+r^2(d\theta^2+\sin^2\theta d\varphi^2),
\end{equation}
where 
\begin{equation}
	f(r)=1-\frac{2M}{r}\left[1-\tanh\left(\frac{Q^2}{2Mr}\right)\right]-\frac{\Lambda}{3}r^2.
\end{equation}
The parameter $M$ serves as the black hole mass and henceforth we set $M=1$ throughout this paper, so we have two free parameters $(Q,\Lambda)$ left. When the values of charge $Q$ and cosmological constant $\Lambda$ are properly chosen, the black hole admits three horizons, i.e. the inner Cauchy horizon  $r_-$, event horizon $r_+$ and cosmological horizon $r_c$. All the parameter combinations $(\Lambda,Q)$ which ensure the presence of three horizons constitute the black hole parameter space, whose boundary can be found by requiring $f(r)=0$ and $f'(r)=0$. Under this condition, the boundary curves of the parameter space can be expressed by parametric equation in terms of parameter $x=Q^2/2r>0$, 
\begin{equation}
\begin{aligned}
&\Lambda(x)=\frac{3[1-x(1+\tanh x)]}{(1-\tanh x)^2[3-x(1+\tanh x)]^3},\\
&Q(x)=\sqrt{2 x(1-\tanh x)[3-x(1+\tanh x)]}.
\end{aligned}
\end{equation}
It is clear that $1-\tanh(x)>0$, and note that  $Q>0$ as well as $\Lambda>0$, so we must have $3-x(1+\tanh x)>0$ and $1-x(1+\tanh x)>0$, which directly restrict the value of $x$ in the range $0<x\lesssim0.639232$. Our analysis finally leads us to the black hole parameter space shown in Fig.~\ref{fig0}.
%%%%%%%%%%%%%%%%%%%%%%%%%%%%%%%
%%%%%%%%%%%%%%%%%%%%%%%%%%%%%%%

We consider imposing a neutral massive  scalar perturbation on this black hole spacetime. The propagation of the massive scalar field is governed by Klein-Gordon equation $(D_{\alpha}D^{\alpha}-\mu^2)\psi=0$, which reduces to
%%%%%%%%%%%%
\begin{equation}
\frac{1}{\sqrt{-g}}\partial_{\alpha}(\sqrt{-g}g^{\alpha\beta}\partial_{\beta}\psi)-\mu^2\psi=0,
\end{equation}
%%%%%%%%%%%%%
where $\mu$ is the mass of scalar field. Decomposing the  scalar field $\psi(t,r,\theta,\phi)$ in terms of spherical harmonic function $Y_{lm}(\theta,\phi)$
\begin{equation}
\psi(t,r,\theta,\phi)=\sum_{l,m}e^{-i\omega t}\frac{\phi(r)}{r}Y_{lm}(\theta,\phi),
\end{equation}
and then we introduce the tortoise coordinate $x$
%%%%%%%%%%
\begin{equation}
dx=\frac{dr}{f(r)}.
\end{equation}
%%%%%%%%%
In the tortoise coordinate, the radial part of Klein-Gordon equation can be greatly simplified to Schr\"{o}dinger-like master equation
%%%%%%
\begin{equation}
\frac{d^2\phi(r)}{dx^2}+\left(\omega^2-V(r)\right)\phi(r) = 0,\label{mastereq}
\end{equation}
%%%%%%
where $V(r)$ is the effective potential
%%%%%
\begin{equation}\label{eq7}
V(r)=f(r)\left(\frac{l(l+1)}{r^2}+\frac{f'(r)}{r}+\mu^2\right).
\end{equation}
%%%%

The calculation of QNF serves as the core task in this work since the QNF will not only  be used to reflect the black hole dynamical properties, but also provide a  test to the validity of SCC.   The spectra of QNMs is related to solving the  eigenvalue problem defined by master equation Eq.~\eqref{mastereq} subjected to the following boundary condition for asymptotically dS and flat spacetime
%%%%%%%%%%%%%
\begin{equation}
\phi \sim
\begin{cases}
   e^{-i\omega x}, & x \to -\infty, \\
   e^{+i\omega x}, &  x \to +\infty,
\end{cases}
\label{master_bc0}
\end{equation}
%%%%%%%%%%%%
where $x \to -\infty$ and $x \to +\infty$ correspond to $r\to r_+$ and $r\to r_c$, respectively. This boundary condition means that,   only the ingoing scalar waves 
are allowed on the event horizon, and on the cosmological horizon we require pure outgoing waves. The QNF lives in complex plane  $\omega=\omega_R+i\omega_I$, with the real part $\omega_R$ standing for oscillation frequency and the imaginary part $\omega_I$ being decay rate. 

\subsection{Numerical methods}

In the search of  QNF, it is inevitable to resort to numerical methods. We will employ three different methods to calculate QNF, including Asymptotic Iteration Method (AIM) \cite{Cho:2011sf},  WKB approximation method  improved by Pade approximants, and Pseudospectral Method (PSM) which has been integrated into a code package  provided by \cite{Jansen:2017oag}. 

\subsubsection{AIM}
To employ  AIM, we need to start from the radial part of Klein-Gordon equation in areal coordinate,
%%%%%%%%%
\begin{equation}
	f(r)f'(r)\phi'(r)+f^2(r)\phi''(r)+(\omega^2-V(r))\phi(r)=0.\label{eq2}
\end{equation}
%%%%%%%%
Then we introduce a new coordinate  $\xi$ defined by
\begin{equation}
	\xi=\frac{1}{r},
\end{equation}
In this new coordinate,  the asymptotical behavior, or the  boundary condition in Eq.~\eqref{master_bc0} of $\phi(r)$ is imposed by  reformulating  $\phi(r)$ in terms of $\xi$ as follows,
\begin{equation}
\phi(\xi)=(\xi_+-\xi)^{-\frac{i\omega}{2\kappa_{+}}}(\xi-\xi_c)^{-\frac{i\omega}{2\kappa_{c}}}\chi(\xi),\label{eq1}
\end{equation}
where $\xi_+ =r_+^{-1}$, $\xi_c=r_c^{-1}$, and $\kappa_+=\frac{|f'(r_+)|}{2}$, $\kappa_c=\frac{|f'(r_c)|}{2}$ are surface gravity of event horizon and cosmological horizon, respectively. After substituting Eq.~\eqref{eq1}  into Eq.~\eqref{eq2}, we arrive at following standard form of  differential equation adopted by AIM, 
\begin{equation}
\chi''(\xi)=\lambda_{0}(\xi) \chi'(\xi)+s_{0}(\xi)\chi(\xi).
\end{equation}
The black hole parameters and the frequency  $\omega$ of scalar field are included in functions  $\lambda_{0}(\xi)$ and $s_{0}(\xi)$ which  are not given here  since  their    expressions are tedious. The more detailed introduction of AIM can be found in \cite{Cho:2011sf}.

\subsubsection{WKB method}
The second method we introduce is the WKB approximation method. For spherically symmetric background, a closed form of QNF  can be given by WKB formula \cite{Konoplya:2019hlu},
%%%%%%%%
\begin{equation}
\begin{aligned}
\omega^2&=V_0+A_2\left(\mathcal{K}^2\right)+A_4\left(\mathcal{K}^2\right)+A_6\left(\mathcal{K}^2\right)+\ldots\\
&-i \mathcal{K} \sqrt{-2 V_2}\left(1+A_3\left(\mathcal{K}^2\right)+A_5\left(\mathcal{K}^2\right)+A_7\left(\mathcal{K}^2\right) \ldots\right),
\end{aligned}
\end{equation}
%%%%%%%%%
where $V_0$ is the effective potential peak value $V_0=V(x_0)$, $x_0$ represents the location of this peak. $V_2$ stands for the value of second order derivative of $V(x)$ with respect to tortoise coordinate $x$ at the potential peak $x_0$.  We  denote the value of  $m$th  order derivative of $V(x)$  at $x_0$ as $V_{m}$, i.e.
%%%%%%%%%%%
\begin{equation}
V_{m}=\left.\frac{d^mV(x)}{dx^m}\right|_{x_0},\quad m\geq2.
\end{equation}
%%%%%%%%%%%%%
 In the WKB formula, $A_{k}(\mathcal{K}^2)$ are polynomials of $V_2,V_3,\ldots V_{2k}$, and each $A_{k}(\mathcal{K}^2)$ should be considered as the $k$th order corrections to the eikonal formula,
%%%%%%%%%%%%%
\begin{equation}
\mathcal{K}=i\frac{\omega^2-V_0}{\sqrt{-2V_2}}.
\end{equation}
%%%%%%%%%%%%
The boundary condition of QNMs constrains  $\mathcal{K}$ to the form
\begin{equation}
\mathcal{K}=n+\frac{1}{2},\quad n\in\mathbb{N},	
\end{equation}
%%%%%%%%%%%%%
where $n$ stands for the overtone number of QNMs. It has been found that the accuracy of WKB method can be improved by Pade approximants \cite{Matyjasek:2017psv}. This optimized  approach is established  from a  definition of a polynomial $P_k(\epsilon)$ \cite{Konoplya:2019hlu},
%%%%%%%%%
\begin{equation}
\begin{aligned}
P_k(\epsilon)&=V_0+A_2\left(\mathcal{K}^2\right)\epsilon^2+A_4\left(\mathcal{K}^2\right)\epsilon^4+A_6\left(\mathcal{K}^2\right)\epsilon^6+\ldots\\
&-i \mathcal{K} \sqrt{-2 V_2}\left(\epsilon+A_3\left(\mathcal{K}^2\right)\epsilon^3+A_5\left(\mathcal{K}^2\right)\epsilon^5+A_7\left(\mathcal{K}^2\right)\epsilon^7 \ldots\right),
\end{aligned}
\end{equation}
%%%%%%%%%%%%
where the polynomial order $k$ is the same as the order of WKB formula. When $\epsilon=1$, one can get 
\begin{equation}
\omega^2=P_k(1).	
\end{equation}
%%%%%%%%%%%%%%%%
One can get more details of WKB method and how the Pade approximants is incorporated  into WKB by referring to \cite{Matyjasek:2017psv,Konoplya:2019hlu} and references therein.

\subsubsection{PSM}

The name of this method suggests that basic notion of this approach is based on  pseudospectral method. As a kind of way of solving differential equation,  it discretizes  the differential equation by replacing  a continuous variable with a discrete set of points called collocation points. Ref. \cite{Jansen:2017oag} offers  a comprehensive understanding of this method, and it also provides a code package which will be used  in the following calculations of QNMs.

\section{Spectra of scalar Quasinormal Modes} \label{sec3}

In this section, we offer the analysis of spectra  of the scalar QNMs in ABGB-dS black hole spacetime. We should point out that the QNF of  massless scalar perturbations of ABGB-dS black hole has been studied in \cite{Fernando:2015fha} by WKB method, here we complement the QNF calculations by including the massive scalar perturbations and using more accurate numerical methods. Our results of QNF are obtained by AIM, PSM and  WKB in order to ensure the reliability of the numerical calculations. Note that we have set $M=1$, which means that all the physical quantities are measured in units of $M^a$. To be specific, the $t,r,Q$ are measured in units of $M$, $\Lambda$ in units of $M^{-2}$, $\omega$ and $\mu$ in units of $M^{-1}$. 

In Table~\ref{tab1}, we show the fundamental QNF with fixed $\Lambda=0.1, Q=0.5$ for various scalar mass $\mu$ and angular number $l$. The comparisons of the QNF among AIM, PSM and WKB directly convince us that we have obtained correct spectra of QNMs due to the excellent agreement between the results, especially for higher $l$ modes as a consequence of the fact that WKB works better for large angular number.  For the QNF with different scalar mass, we can see that all the real parts of QNF increase with the angular number, as a reasonable expectation that larger $l$ correspond to higher oscillation frequency. While for the imaginary parts of QNF, the effects of scalar mass will make QNMs behave differently.  For the massless and  massive  QNMs with small mass values (related to $\mu=0,0.1$ in the table), the imaginary part grows (smaller magnitude) with the angular number, which indicates that the dominant  QNF comes from the QNMs at large $l$ limit.  For the larger scalar mass ($\mu=0.3, 0.5$), the imaginary part decreases (greater magnitude) when we improve the  angular number. Therefore we have $l=0$ the most fundamental modes in larger scalar mass cases. Apparently, the distinctions of impacts of angular number  on QNF are attributed to the scalar mass. One can observe that, for QNF with larger mass, both the real and imaginary part of QNF will have higher values, which means that the heavier scalar QNMs oscillate more rapidly and  have lower damping rate.

%%%%%%%%%%%%%%%%%%%%%%%%%%%%%%%
\begin{table}[!htbp]
\centering
\caption*{$\Lambda=0.1, Q=0.5$} 
\resizebox{\textwidth}{!}
{
        \begin{tabular}{ccccccc}
    \hline\hline
    $\mu$ &Method&  $l=0$                    &$l=1$                     & $l=3$                & $l=5$                      & $l=10$       \\
    \hlineB{2}
        &PSM&     $ 0.0297268-0.0700588i $  & $ 0.115393-0.0427651i $ & $0.291886-0.0413015i   $ & $ 0.463129-0.0411367i$    & $0.888306-0.0410575i  $\\
        \cline{2-7}
  $0$   &AIM&     $ 0.0297268 -0.0700588i$  & $ 0.115393-0.0427651i $ & $0.291886 -0.0413015i  $ & $ 0.463129 -0.0411367i$   & $0.888306-0.0410575i  $\\
        \cline{2-7}
        &WKB&     $ 0.0232785 -0.0639364i$  & $ 0.115358 -0.0428801i $ & $ 0.291886 -0.0413024i$ & $0.463129  -0.0411368i $   & $0.888306 -0.0410575i$\\
    \hlineB{2}
        &PSM&     $ 0 -0.00839543i$       & $ 0.117866-0.0424253i  $ & $0.292865-0.0412586i  $ & $0.463747-0.0411201i $     & $ 0.888628-0.0410531i $\\
        \cline{2-7}
  $0.1$   &AIM&   $ 0 - 0.00839543i $       & $ 0.117866 -0.0424253i $ & $0.292865 -0.0412586i  $ & $0.463747 -0.0411201i$   & $0.888628  -0.0410530i  $\\
        \cline{2-7}
        &WKB&     $ 0.0268185 -0.0607713i$  & $ 0.117825 -0.0425127i $ & $0.292865  -0.0412586i $ & $0.463746  -0.0411202i $   & $0.888628  - 0.0410531i$\\
    \hlineB{2}
       &PSM&      $0.0655197-0.0398826i $ & $ 0.136293-0.0406433i $  & $ 0.300614-0.0409406i $   & $ 0.468666-0.0409915i$       & $0.891202-0.0410178i  $\\
        \cline{2-7}
  $0.3$   &AIM&   $ 0.0655197-0.0398826i $  & $ 0.136293 -0.0406433i  $ & $0.300614  -0.0409405i $ & $ 0.468666 -0.0409915i $  & $0.891202  -0.0410179i  $\\
        \cline{2-7}
        &WKB&     $ 0.0660754 -0.0397961i$  & $ 0.136301 -0.0406336i $ & $0.300614  -0.0409408i $ & $ 0.468666 - 0.0409916i$   & $0.891202  - 0.0410178i$\\
    \hlineB{2}
       &PSM&      $0.118485-0.0384445i $   & $ 0.167651-0.0392526i  $ & $0.315659-0.0404201i  $  & $0.478389-0.0407545i $       & $0.896334-0.0409488i  $\\
        \cline{2-7}
  $0.5$  &AIM&   $ 0.118485 -0.0384445i $  & $ 0.167651 -0.0392526i  $ & $0.315659  -0.0404201i$ & $0.478389  -0.0407545i$     & $ 0.896334 - 0.0409489i $\\
        \cline{2-7}
        &WKB&     $ 0.11833 -0.0384774i$    & $ 0.167634 -0.0392564i $ & $ 0.315660 -0.040420i $ & $ 0.478389 - 0.0407546i$    & $ 0.896334 -0.0409488i $\\
        \hline\hline
\end{tabular}
}
\caption{The fundamental QNF obtained by PSM, AIM and WKB  at $\Lambda=0.1, Q=0.5$ for various mass values $\mu$ and angular number $l$. A purely imaginary QNF at $\mu=0.1, l=0$ is found by PSM and AIM while missed by WKB.}\label{tab1}
\end{table}
%%%%%%%%%%%%%%%%%%%%%%%%%%%%%%%

\begin{figure}[thbp]
\centering
\includegraphics[height=2.4in,width=3.2in]{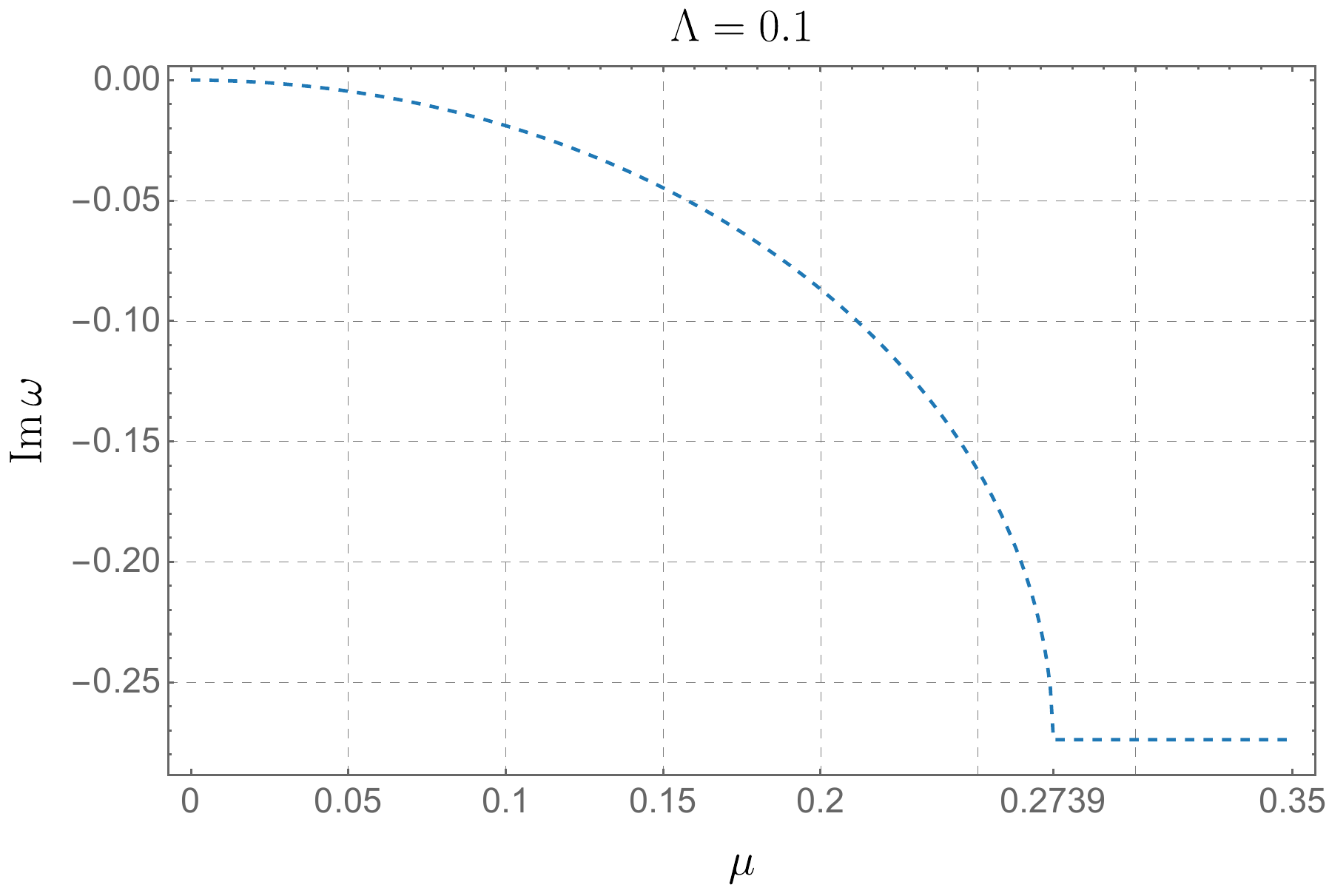}
\includegraphics[height=2.4in,width=3.2in]{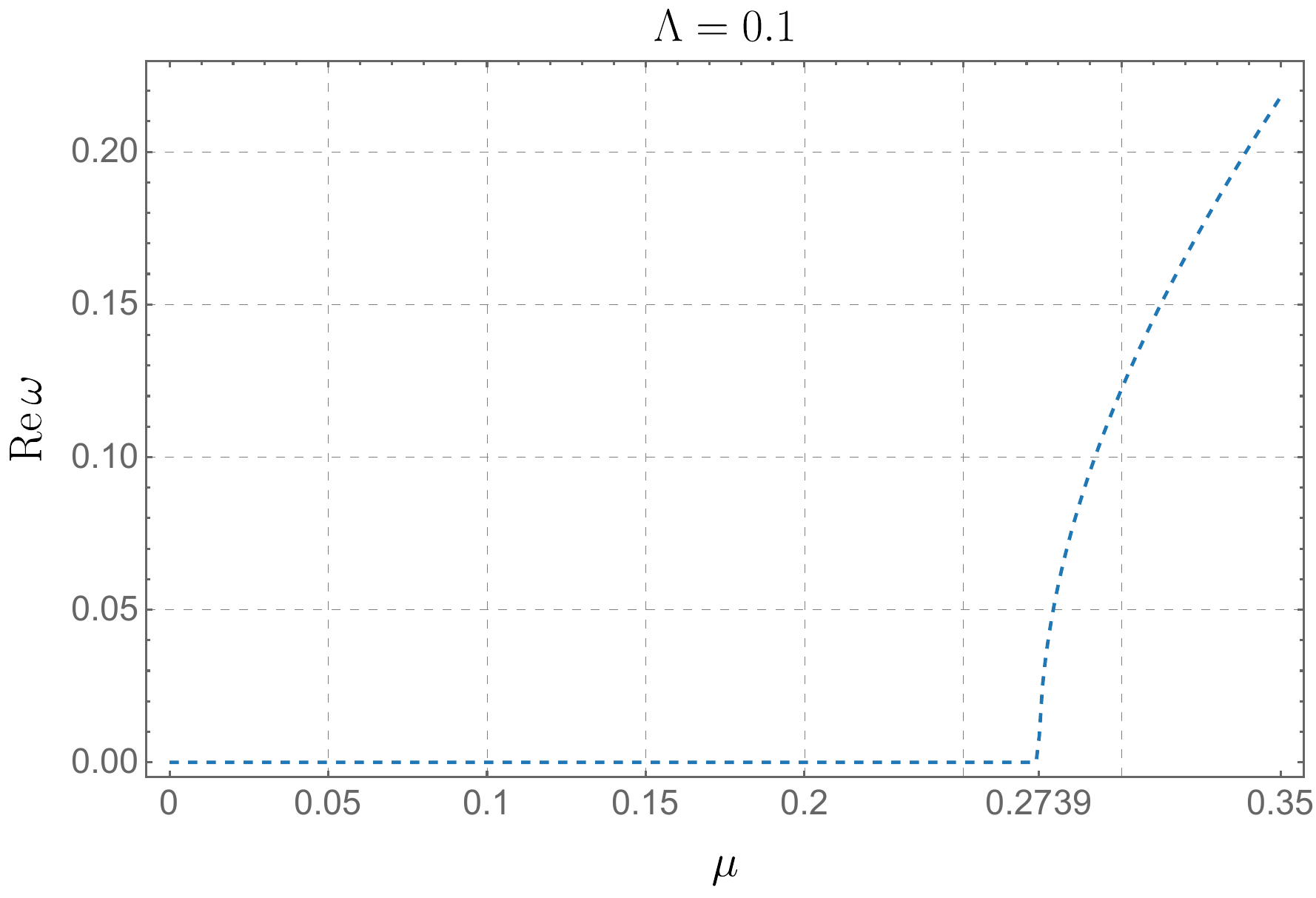}
\caption{The dependence of massive scalar QNF in pure dS spacetime on the scalar mass for $l=n=0$. The critical scalar field mass is located at $\mu_{\rm{cri}}\simeq0.2739$.\label{fig4}}
\end{figure}

One must have noticed a special QNF which is purely imaginary at $\mu=0.1$, $l=0$. This frequency is captured by both PSM and AIM, providing an independent numerical confirmation of this mode, whereas the WKB approximation missed it since the characteristics of WKB prevent it from calculating purely imaginary frequencies of QNMs.  Actually, this pure damping  mode without oscillation is traced back to the constant mode $\phi_0\in \mathbb{C} $ (or ``zero-mode'', as its  $\omega=\omega_R+i\omega_I=0$, present for $l=0, n=0$) which is related to massless scalar perturbations  and lives in some asymptotically dS spacetime. In fact, this massless zero-mode is a member of  purely imaginary QNMs found in pure dS spacetime (hence they are called pure dS modes) \cite{Du:2004jt,Lopez-Ortega:2006aal} and RN-dS spacetime \cite{Cardoso:2017soq}, and a nonvanishing scalar mass can make this kind of QNMs become complex if the mass value is above some critical value. This fact is clearly expressed by a closed formula of massive scalar QNF in pure  dS spacetime \cite{Lopez-Ortega:2006aal}. With the spacetime dimension $D=4$, this formula is given by
%%%%%%%%%%%%%
\begin{equation}
\omega =
\begin{cases}
   -i\kappa\left(l+2n+\frac{3}{2}-\sqrt{\frac{9}{4}-\frac{\mu^2}{\kappa^2}}\right), \quad \frac{9}{4}-\frac{\mu^2}{\kappa^2}>0 \\[3ex]
    -i\kappa\left(l+2n+\frac{3}{2}+i\sqrt{\frac{\mu^2}{\kappa^2}-\frac{9}{4}}\right), \quad \frac{9}{4}-\frac{\mu^2}{\kappa^2}<0
\end{cases}
\label{massive puredsmode}
\end{equation}
%%%%%%%%%%%%
where $\kappa$ is the surface gravity of cosmological horizon of the pure dS spacetime, and $n$ is the overtone number. From Eq.~\eqref{massive puredsmode}, it is easy to see that purely imaginary modes emerge only for $\mu\leq\mu_{\rm{cri}}=3/2\kappa$, and the ``zero-mode'' is present when $\mu=0, l=n=0$. The dependence  of this pure dS modes with $l=n=0$ on the scalar mass is shown in Fig.~\ref{fig4}, which shows that either the imaginary part or the real part of the mode remains a constant.

%%%%%%%%%%%%%%%%%%%%%%%%%%%%%%%
%%%%%%%%%%%%%%%%%%%%%%%%%%%%%%%

\begin{figure}[thbp]
\centering
\includegraphics[height=2.4in,width=3.2in]{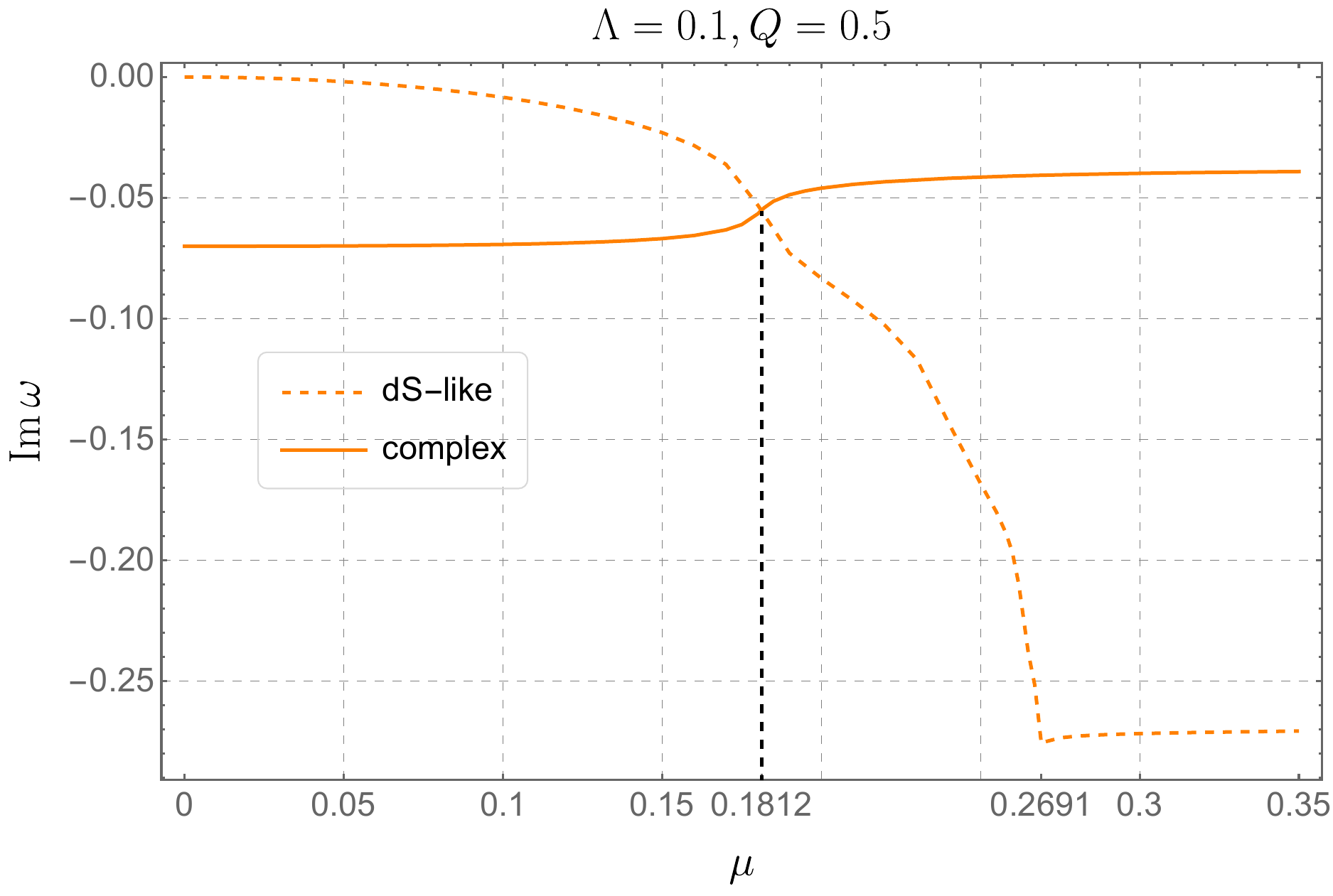}
\includegraphics[height=2.4in,width=3.2in]{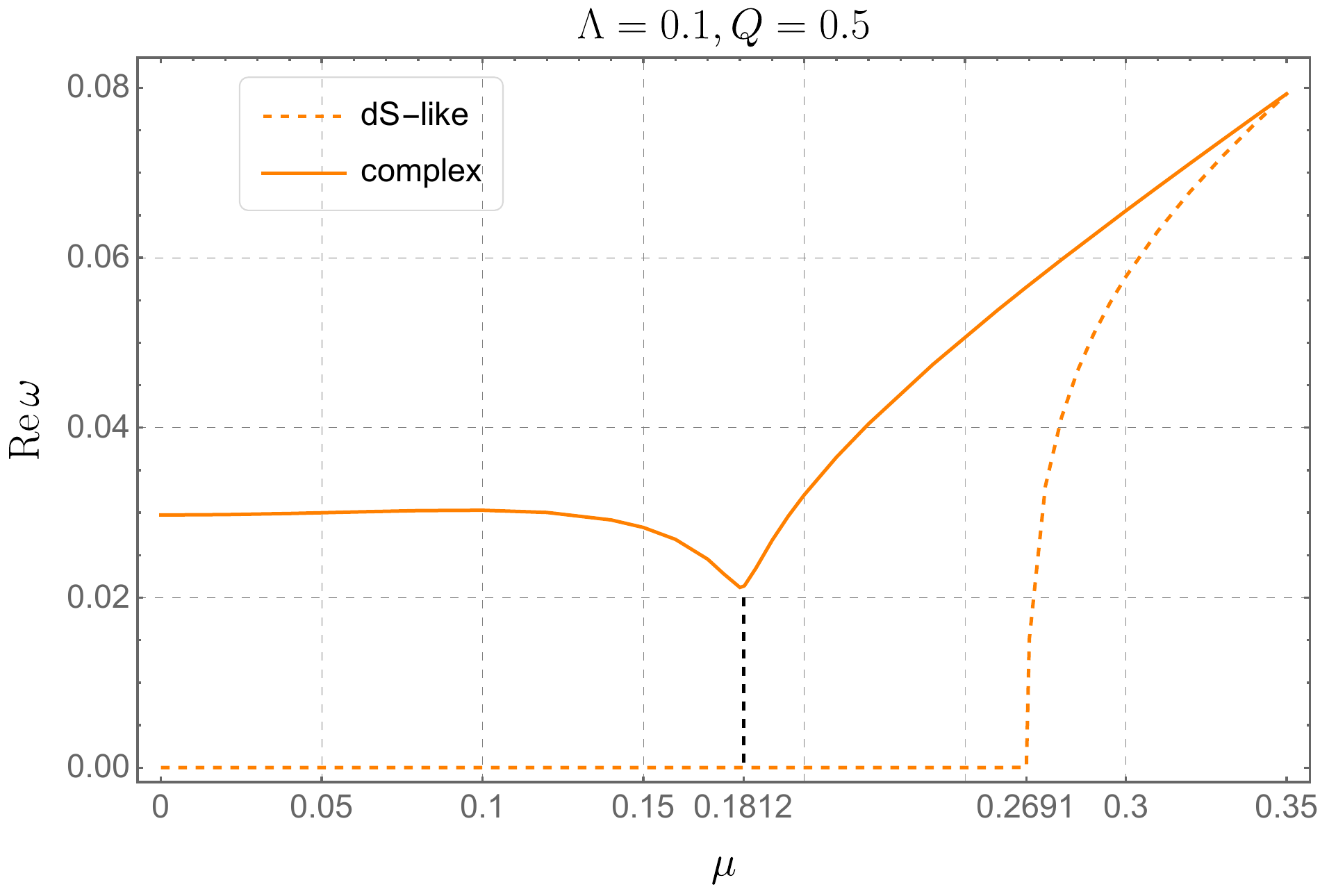}
\caption{The behaviors of imaginary and real part of fundamental  QNF obtained by PSM of two classes of QNMs with $l=0$ under the change of scalar mass $\mu$. The dashed orange curve denotes the dS-like mode, and solid orange curve denotes the complex mode, i.e. the real part remains non-vanishing.\label{fig1}}
\end{figure}

Interestingly,  we find that a certain branch of the massive scalar QNMs in  ABGB-dS spacetime inherit similar features from the massive scalar QNMs in pure dS spacetime (further discussions will unfold in next section). We refer to this branch of QNMs as dS-like mode demonstrated in Fig.~\ref{fig1}.  In this figure, we demonstrate the imaginary and real part  of fundamental  QNF of dS-like mode and ordinary complex mode with $l=0$. The dS-like mode gets its name for its similar behavior under the change of scalar mass to the pure dS modes shown in Fig.~\ref{fig4}. When  scalar field starts to possess  mass, the zero-mode no longer exists as it deviates  from $\omega=0$, as shown in Fig.~\ref{fig1}.  We see that the dS-like mode originates from $\omega=0$ of zero-mode at limit $\mu\to0$, and it serves as dominant mode versus  complex mode in the mass region $\mu\lesssim 0.1812$, at which the real part of complex mode develops a sharp minimum. The purely imaginary segment of dS-like mode exists in the mass region $\mu\leq\mu_{\rm{cri}}\simeq0.2691$, which is a bit smaller than the critical mass $\mu_{\rm{cri}}\simeq0.2739$ in pure dS spacetime, and the imaginary part of this mode quickly decreases with the scalar mass. In contrast, the imaginary part of complex mode changes mildly when changing scalar mass, and it exhibits its maximum rate of change with respect to the scalar mass  at approximately $\mu=0.1812$, i.e. corresponding to the crossing  of the damping rate of the two modes. For $\mu>0.2691$, the imaginary part of dS-like mode shows a weak dependence on scalar mass, while its real part grows quickly with the mass, just like the property also found in pure dS spacetime.  The fact that the  purely imaginary segment of dS-like mode only dominates  in small scalar mass region explains the occurrence  of pure decay QNF at $\mu=0.1$ in Table~\ref{tab1}. On the other hand, we list some numerical results of QNF of dS-like mode for several scalar mass values in Table~\ref{tab7}, which demonstrates a great  agreement between the results from  PSM and AIM, and provides an independent numerical confirmation of this mode. More importantly, the frequency continuously approaches $\omega=0$ as $\mu$ decreases, providing direct numerical evidence that the dS-like mode is continuously connected to the massless zero-mode in the limit $\mu\to0$. 

%%%%%%%%%%%%%%%%%%%%%%%%%%%%%%%
%%%%%%%%%%%%%%%%%%%%%%%%%%%%%%%

%%%%%%%%%%%%%%%%%%%%%%%%%%%%%%%
\begin{table}[!htbp]
\centering
\setlength{\tabcolsep}{6pt}
\caption*{$\Lambda=0.1, Q=0.5$} 
%\resizebox{\textwidth}{!}
{
        \begin{tabular}{cccccc}
    \hline\hline
    Method  &  $\mu=0.001$                     &$\mu=0.05$                     & $\mu=0.1$             & $\mu=0.15$                 & $\mu=0.2$       \\
    \hlineB{2}
      PSM   &  $0-7.58481\times 10^{-7}i$   & $0-0.00193978i$                  & $0-0.00839543i$        & $0-0.0230291i$          & $0-0.0833359i$   \\
      \hline
      AIM  &   $0-7.58481\times 10^{-7}i$   & $0-0.00193978i$                  & $0-0.00839543i$        & $0-0.0230291i $         & $0-0.0833359i$  \\
        \hline\hline
\end{tabular}
}
\caption{The fundamental QNF of dS-like mode with $l=0$ obtained by PSM, AIM at  $\Lambda=0.1, Q=0.5$ for several scalar mass values.}\label{tab7}
\end{table}
%%%%%%%%%%%%%%%%%%%%%%%%%%%%%%%

%%%%%%%%%%%%%%%%%%%%%%%%%%%%%%%
\begin{table}[!htbp]
\centering
\caption*{$\Lambda=0.1, \mu=0.2$} 
\resizebox{\textwidth}{!}
{
        \begin{tabular}{ccccccc}
    \hline\hline
    $Q$ &Method&  $l=0$                     &$l=1$                     & $l=3$                    & $l=5$                       & $l=10$       \\
    \hlineB{2}
           &PSM&  $0.0232554-0.0327255i$    & $0.0909906-0.0311631i$   & $0.215799-0.0309507i$   &  $0.339808-0.0309205$        & $0.649375-0.0309054i$   \\
        \cline{2-7}
    $0.1$ &AIM&   $0.0232554 -0.0327255i$   & $0.0909906 -0.0311631i $ & $0.215799  -0.0309507i $ & $0.339808  -0.0309205i $   & $ 0.649375 - 0.0309054i $\\
        \cline{2-7}
        &WKB&     $0.0234288 -0.0343879i$   & $0.0909714 -0.0311798i $ & $0.215798  -0.0309515i $ & $0.339808  -0.0309206i $   & $ 0.649375 - 0.0309054i $\\
    \hlineB{2}
          &PSM&   $0.0261886-0.0371645i$    & $0.102647-0.0348535i$   &  $0.243245-0.0345650i$   &  $0.382982-0.0345245i$      & $0.731836-0.0345043i$   \\
        \cline{2-7}
    $0.3$ &AIM&   $0.0261886 -0.0371645i$   & $0.102647  -0.0348535i $ & $0.243245  -0.0345650i $ & $0.382982  -0.0345245i $   & $ 0.731836 - 0.0345043i $\\
        \cline{2-7}
        &WKB&     $0.0268012 -0.0390231i$   & $0.102627  -0.0348756i $ & $0.243245  -0.0345656i $ & $0.382982  -0.0345245i $   & $ 0.731836 - 0.0345043i $\\
    \hlineB{2}
          &PSM&   $0.0320850-0.0460065i$    & $0.125047-0.0415940i$   &  $0.295790-0.0411342i$   &   $0.465596-0.0410711i$     & $0.889594-0.0410398i$   \\
        \cline{2-7}
    $0.5$ &AIM&   $0.0320850 -0.0460065i$   & $0.125047  -0.0415940i $ & $0.295790  -0.0411342i  $ & $0.465596  -0.0410711i $  & $ 0.889594 - 0.0410398i $\\
        \cline{2-7}
        &WKB&     $0.0339030 -0.0475451i$   & $0.125022  -0.0416217i $ & $0.295790  -0.0411347i $ & $0.465596  -0.0410711i $    & $ 0.889594 - 0.0410398i $\\
    \hlineB{2}
         &PSM&   $0.0523529-0.0674296i$    & $0.181671-0.0557658i$   &  $0.427424-0.0548588i$   &  $0.672343-0.0547396i$       & $1.28419-0.0546810i$   \\
        \cline{2-7}
    $0.8$ &AIM&   $0.0523529 -0.0674296i$   & $0.181671  -0.0557658i $ & $0.427424  -0.0548588i $ & $0.672343  -0.0547396i  $  & $ 1.28419 - 0.0546810i $\\
        \cline{2-7}
        &WKB&     $0.0548607 -0.0652054i$   & $0.181668  -0.055785i  $ & $0.427423  -0.0548586i $ & $0.672343  -0.0547396i $    & $ 1.28419 - 0.0546810i $\\
        \hline\hline
\end{tabular}
}
\caption{The fundamental QNF obtained by PSM, AIM and WKB  at $\Lambda=0.1, \mu=0.2$ for various charge values $Q$ and angular number $l$.}\label{tab2}
\end{table}
%%%%%%%%%%%%%%%%%%%%%%%%%%%%%%%

In Table~\ref{tab2}, we show the QNF with fixed $\Lambda=0.1, \mu=0.2$ for various values of charge $Q$ in order to reflect the influences of charge on QNF. The data in this table show that, for a fixed $Q$, the dominant modes correspond to larger angular number $l$. When increasing charge value, we find that the real part increases while the imaginary part decreases, therefore the QNMs will oscillate more rapidly and decay faster for a  larger value of $Q$. In Table~\ref{tab3} we demonstrate the influences of cosmological constant $\Lambda$ on QNF. It is found that a larger $\Lambda$ will reduce the oscillation frequency of QNMs, while increasing the imaginary part leading to a slower  decaying rate. So in this sense, we see that  the cosmological constant and charge affect the QNF in opposite ways. When considering the angular number, we note that it changes the imaginary part of  QNF in different manner depending on the value of $\Lambda$. For a small $\Lambda=0.05$, larger angular number lowers  the imaginary part and thus the  $l=0$ mode is the dominant one. However,  this is not the case for bigger $\Lambda$ under which the dominant mode corresponds to large $l$.

%%%%%%%%%%%%%%%%%%%%%%%%%%%%%%%
\begin{table}[!htbp]
\centering
\caption*{$Q=0.8,\mu=0.2$} 
\resizebox{\textwidth}{!}
{
        \begin{tabular}{ccccccc}
    \hline\hline
    $\Lambda$ &Method&   $l=0$                    &$l=1$                    & $l=3$                  & $l=5$                       & $l=10$       \\
    \hlineB{2}
              &PSM&     $0.111099 -0.0755964i$   & $0.274815 -0.0790031i$  & $0.625990 -0.0793861i$ & $0.980416 -0.0794350i$      & $1.86860 -0.0794591i$   \\
        \cline{2-7}
    $0.05$ &AIM&         $0.111099 -0.0755964i$   & $0.274815 -0.0790031i $ & $0.625990 -0.0793860i$ & $0.980416 -0.0794350i $     & $1.86860 -0.0794592i $\\
        \cline{2-7}
        &WKB&            $0.106922 -0.0752779i$   & $0.27481 -0.07901i $    & $0.625990 -0.0793861i $ & $0.980416 -0.0794350i $    & $1.86860 -0.0794591i $\\
    \hlineB{2}
         &PSM&           $0.0763594 -0.0750709i$  & $0.222072 -0.0667687i$  & $0.515410 -0.0659062i$  & $0.809319 -0.0657929i$     & $1.54448 -0.0657373i$   \\
        \cline{2-7}
    $0.08$ &AIM&         $0.0763594 -0.0750709i$  & $0.222072 -0.0667687i $ & $0.515410 -0.0659062i $ & $0.809319 -0.0657929i $    & $1.54448 -0.0657373i $\\
        \cline{2-7}
        &WKB&            $0.0774283 -0.0743575i$  & $0.222063 -0.0667785i $ & $0.515409 -0.0659061i $ & $0.809319 -0.0657929i $    & $1.54448 -0.0657373i $\\
     \hlineB{2}
          &PSM&         $0.0523529 -0.0674296i$  & $0.181671 -0.0557658i$  & $0.427424 -0.0548588$   & $0.672343 -0.0547396i$     & $1.28419 -0.0546810i$   \\
        \cline{2-7}
    $0.1$ &AIM&          $0.0523529 -0.0674296i$  & $0.181671 -0.0557658i $ & $0.427424 -0.0548588i $ & $0.672343 -0.0547396i $    & $1.28419 -0.0546810i $\\
        \cline{2-7}
        &WKB&            $0.0548607 -0.0652054i$  & $0.181668 -0.0557850i $ & $0.427423 -0.0548586i $ & $0.672343 -0.0547396i $    & $1.28419 -0.0546810i $\\
     \hlineB{2}
          &PSM&        $0.0272529 -0.0497754i$  & $0.133028 -0.0413553i$  & $0.317334 -0.0408204i$  & $0.500053 -0.0407492i$     & $0.955940 -0.0407141i$   \\
        \cline{2-7}
    $0.12$ &AIM&         $0.0272529 -0.0497754i$  & $0.133028 -0.0413553i $ & $0.317334 -0.0408204i $ & $0.500053 -0.0407492i $    & $0.955940 -0.0407141i $\\
        \cline{2-7}
        &WKB&            $0.0265239 -0.0447199i$  & $0.133006 -0.0413798i $ & $0.317333 -0.0408205i $ & $0.500053 -0.0407492i $    & $0.955940 -0.0407141i $\\
        \hline\hline
\end{tabular}
}
\caption{The fundamental QNF obtained by PSM, AIM and WKB  at $Q=0.8, \mu=0.2$ for various cosmological constant $\Lambda$ and angular number $l$.}\label{tab3}
\end{table}
%%%%%%%%%%%%%%%%%%%%%%%%%%%%%%%

\section{Strong Cosmic Censorship}\label{sec4}

In this section we start to investigate the SCC in ABGB-dS spacetime. We will separately discuss this subject in the case of massless and massive scalar perturbations. We recall that the criterion of the violation of SCC is given by
%%%%%
\begin{equation}
	\beta\equiv \frac{\alpha}{\kappa_-}>\frac{1}{2}, \quad \alpha=-\mathrm{Im}\,\omega,\label{criterion}
\end{equation}  
%%%%%%
where $\kappa_-=|f'(r_-)|/2$ is the surface gravity of Cauchy horizon, $\alpha$ is the spectral gap defined by the imaginary part of the dominant non-zero QNMs. Let us put it another way, for a given set of black hole and scalar field parameters, the SCC will be violated under scalar perturbations  if all the non-zero QNF  $\{\omega_{ln}\}$ satisfy $-\mathrm{Im}\,\omega_{ln}/\kappa_->1/2$, otherwise the SCC is respected. Therefore, our main task in this section is to find out the slowest damping rate of QNMs among $\{\omega_{ln}\}$ and use it to examine the inequality Eq.~\eqref{criterion}. Note that we will use PSM in the following calculations of QNMs spectra.

\subsection{Classifications of QNMs for massless scalar perturbations}

It has been found that the massless  QNMs can be classified into three families, photon sphere (PS) modes, de Sitter (dS) modes and near-extremal (NE) modes in RN-dS spacetime \cite{Cardoso:2017soq}. The same classification of massless QNMs  is expected in ABGB-dS spacetime due to its  similar structure to RN-dS spacetime,   and we will demonstrate  that it is indeed the case. Henceforth, for the convenience of following discussions, we will refer to $-\mathrm{Im}\,\omega/\kappa_-$ obtained by fundamental PS, dS and NE modes to $\beta_{\mathrm{PS}},\beta_{\mathrm{dS}}$ and $\beta_{\mathrm{NE}}$, respectively, meanwhile the dominant one among $\beta_{\mathrm{PS,dS,NE}}$ is represented by $\beta$ as how it is defined.

The first family of QNMs is the PS modes which have non-vanishing real part $\omega_R$. The reason for the name is that this family of QNMs are connected to photon sphere, and the connection is established by WKB approximation of QNF at  eikonal limit ($l\to\infty$). It is well-known that, in the eikonal regime, WKB formula is able to provide accurate QNF. Intriguingly, it is found in \cite{Cardoso:2008bp} that this WKB formula  can be rewritten as 
%%%%%%%%
\begin{equation}
\omega_{\mathrm{eikonal}}=\Omega_cl-i\left(n+\frac{1}{2}\right)|\lambda|,\label{ps formula}
\end{equation}
%%%%%%%%%     
where $\Omega_c$ and $\lambda$ is the coordinate angular velocity and  Lyapunov exponent of unstable circular null geodesics at photon sphere, respectively. Lyapunov exponent is used to measure the instability timescale of unstable circular null geodesics. The dominant QNF of PS modes is given by Eq.~\eqref{ps formula} for overtone number  $n=0$. In our later calculations of $\beta_{\mathrm{PS}}$, we will use the fundamental QNF with $l=10$  as a representative of the dominant QNF from eikonal limit $l\to\infty$, since it can provide a good approximation demonstrated in Table \ref{tab4}  to dominant QNF in PS family.

%%%%%%%%%%%%%%%%%%%%%%%%%%%%%%%
\begin{table}[!htbp]
\centering
%\resizebox{\textwidth}{!}
{
        \begin{tabular}{|c|c|c|c|c|}
    \hline    % after \\: \hline or \cline{col1-col2} \cline{col3-col4} ...
    Method&                  $Q/Q_{\mathrm{max}}=0.99$  &$Q/Q_{\mathrm{max}}=0.992$   & $Q/Q_{\mathrm{max}}=0.995$   & $Q/Q_{\mathrm{max}}=0.999$    \\
    \hline
     WKB($l\to\infty$) &      0.431034                  & 0.496532                    & 0.662088                     & 1.65466    \\
        
        \hline
               
          PSM($l=10$)  &      0.431399                  & 0.496950                    & 0.662643                     & 1.65605   \\

           \hline
           
           \end{tabular}
}
\caption{The comparison of $\beta_{\mathrm{PS}}$ obtained by WKB at eikonal limit and PSM at $l=10$ with $\Lambda=0.1$ for several values of ratio $Q/Q_{\mathrm{max}}$ .}\label{tab4}
\end{table}
%%%%%%%%%%%%%%%%%%%%%%%%%%%%%%%

The second family of the QNMs is the dS modes. The main feature of the dS modes is that they inherit some properties from the pure dS spacetime, which means that the QNF of dS modes are purely imaginary and can be well approximated, especially for dominant dS mode, by analytical formula of spectra for pure dS QNMs \cite{Du:2004jt,Lopez-Ortega:2006aal,Vasy:2007tda}
%%%%%%%%%%%%%%%%
\begin{equation}
\begin{aligned}
	&\omega_{n=0,\mathrm{pure\, dS}}/\kappa_c^{\mathrm{dS}}=-il,\\
	&\omega_{n\neq 0,\mathrm{pure\, dS}}/\kappa_c^{\mathrm{dS}}=-i(l+n+1),\label{dS QNF}
\end{aligned}
\end{equation}
%%%%%%%%%%%%%%%%
where $\kappa_c^{\mathrm{dS}}=\sqrt{\Lambda/3}$ is the surface gravity of cosmological horizon of pure dS spacetime, $n$ is overtone number. When $l=0, n=0$, we get a zero-mode with $\omega=0$, which will be modified by introducing scalar mass. In fact, the introduction of a large enough scalar mass will make the purely imaginary modes become complex \cite{Lopez-Ortega:2006aal}, as what we have previously illustrated in Sec.~\ref{sec3}.  The dominant dS mode corresponds to $l=1,n=0$, and we demonstrate a comparison of this dominant QNF obtained by  Eq.~\eqref{dS QNF} and spectral method in Table~\ref{tab5}. One can observe that the dominant dS modes in ABGB-dS spacetime are well approximated by the pure dS modes in pure dS spacetime and they are  weakly dependent on charge $Q$.

%%%%%%%%%%%%%%%%%%%%%%%%%%%%%%%
\begin{table}[!htbp]
\centering
\caption*{$l=1,n=0$} 
%\resizebox{\textwidth}{!}
{
        \begin{tabular}{|c|c|c|c|c|}
    \hline    % after \\: \hline or \cline{col1-col2} \cline{col3-col4} ...
    Method&                  $Q/Q_{\mathrm{max}}=0.8$  &$Q/Q_{\mathrm{max}}=0.99$   & $Q/Q_{\mathrm{max}}=0.995$   & $Q/Q_{\mathrm{max}}=0.999$    \\
    \hline
     $\omega_{n=0,\mathrm{pure\, dS}}$ &          $-0.182574 i$              &  $-0.182574 i$                    &  $-0.182574 i$             &  $-0.182574 i$    \\
        
        \hline
               
     $\omega_{\mathrm{PSM}}$       &         $-0.182347 i$              &  $-0.182696 i$                    &   $-0.182706 i$             & $-0.182714 i$  \\

           \hline
           
           \end{tabular}
}
\caption{The comparison of QNF of dominant pure dS modes by Eq.~\eqref{dS QNF} and dominant dS modes obtained by PSM  at $\Lambda=0.1$. }\label{tab5}
\end{table}
%%%%%%%%%%%%%%%%%%%%%%%%%%%%%%%

The third family of QNMs is the NE modes. The reason for the name of NE modes is that this family of QNMs manifest  themselves when black holes approach extremality ($r_-\sim r_+$). For a RN-dS black hole at $r_-\to r_+$ limit, the QNF of the NE modes will approach   
%%%%%%%%%%%%
\begin{equation}
	\omega_{\mathrm{NE}}=-i(l+n+1)\kappa_-=-i(l+n+1)\kappa_+,\label{NE QNF}
\end{equation}
%%%%%%%%%%%
which is also purely imaginary and the dominant mode comes from $l=n=0$. Although this formula is derived for RN-dS black hole, in Table~\ref{tab6} we numerically show that it still holds for our ABGB-dS black hole. Apparently, the NE modes will be dominant among all the three families of QNMs at the near extremality, since the value of  $-\mathrm{Im}\,\omega/\kappa_-$ for NE modes is limited, while for the PS and dS modes it  diverges as $\kappa_-\to 0$ in the process of reaching extremality.

%%%%%%%%%%%%%%%%%%%%%%%%%%%%%%%
\begin{table}[!htbp]
\centering
\caption*{$l=n=0$} 
%\resizebox{\textwidth}{!}
{
        \begin{tabular}{|c|c|c|c|c|}
    \hline    % after \\: \hline or \cline{col1-col2} \cline{col3-col4} ...
    NE modes                           &     $Q/Q_{\mathrm{max}}=0.99$  &$Q/Q_{\mathrm{max}}=0.999$   & $Q/Q_{\mathrm{max}}=0.9999$   & $Q/Q_{\mathrm{max}}=0.99999$    \\
    \hline
                   
     $\beta_{\mathrm{NE}}$   &    0.871376  &      0.923424            &  0.973351                     & 0.9915  \\

           \hline
           
           \end{tabular}
}
\caption{The value of $\beta_{\mathrm{NE}}$ for dominant NE modes  at $\Lambda=0.1$ for several high charge ratios.  When we improve  the extremality of black hole by increasing  $Q/Q_{\mathrm{max}}$, $\beta_{\mathrm{NE}}$ will approach unity as described by Eq.~\eqref{NE QNF}.}\label{tab6}
\end{table}
%%%%%%%%%%%%%%%%%%%%%%%%%%%%%%%

\subsection{SCC for massless scalar perturbations}

\begin{figure}[thbp]
\centering
\includegraphics[height=2.4in,width=3.2in]{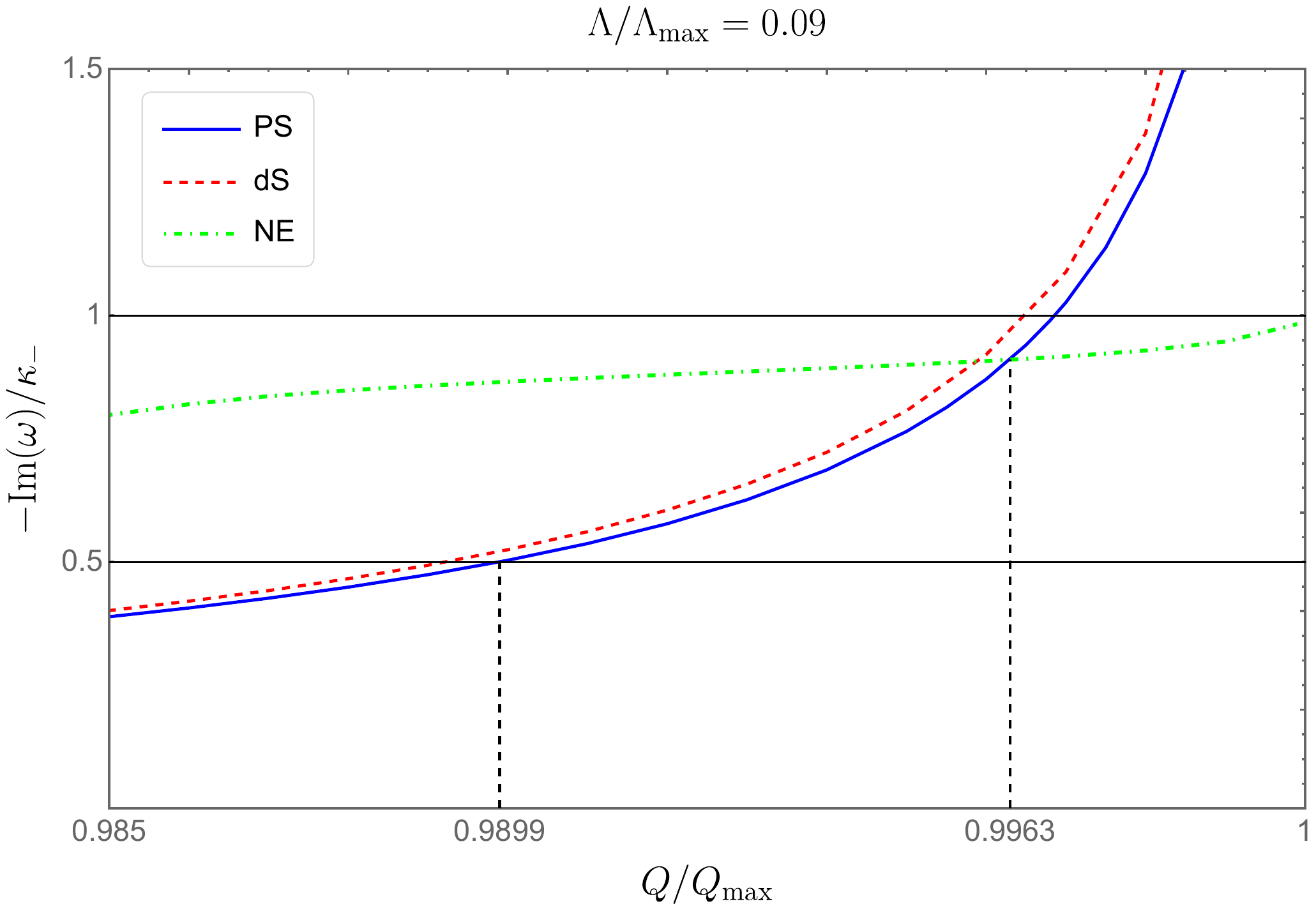}
\includegraphics[height=2.4in,width=3.2in]{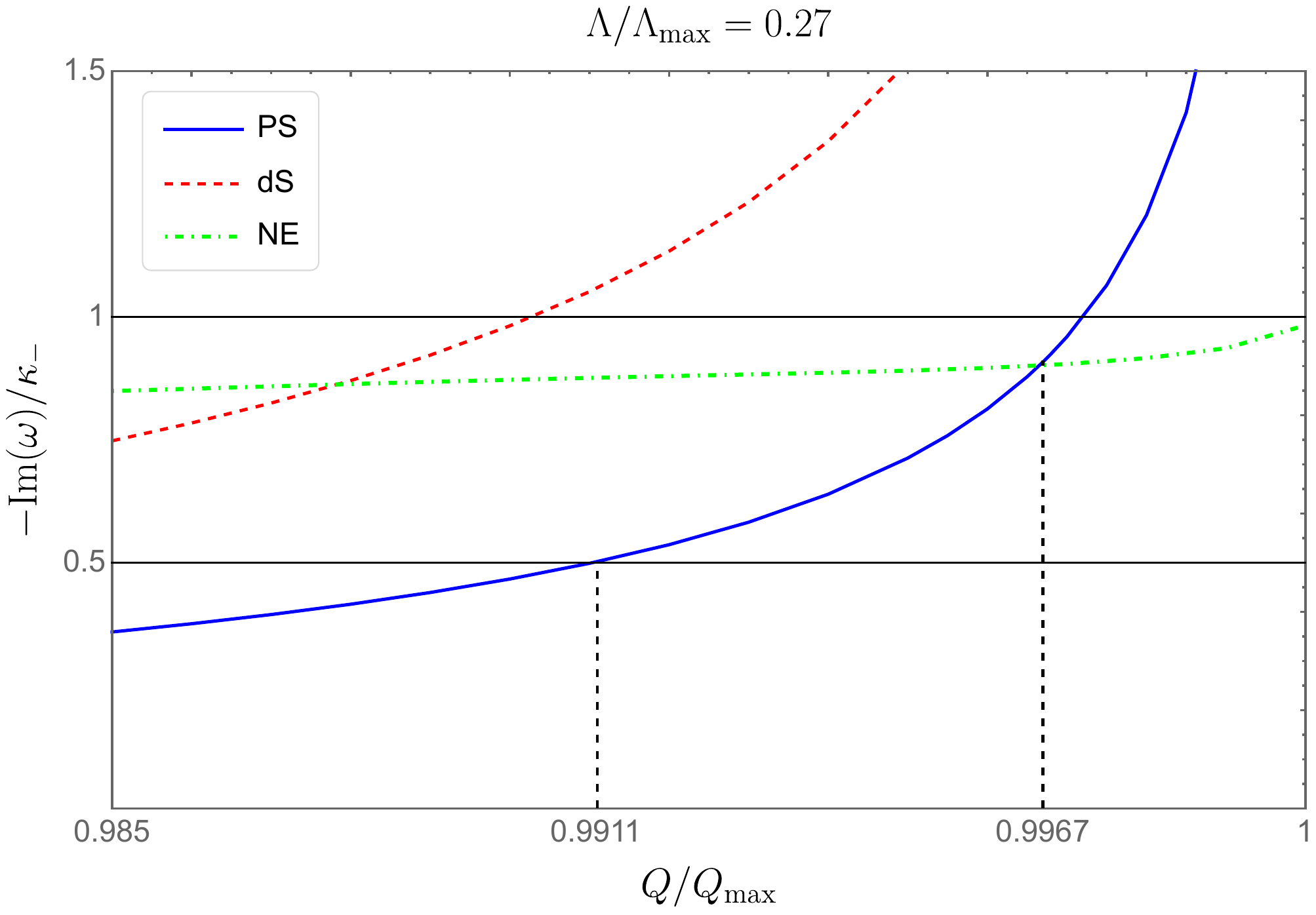}

\includegraphics[height=2.4in,width=3.2in]{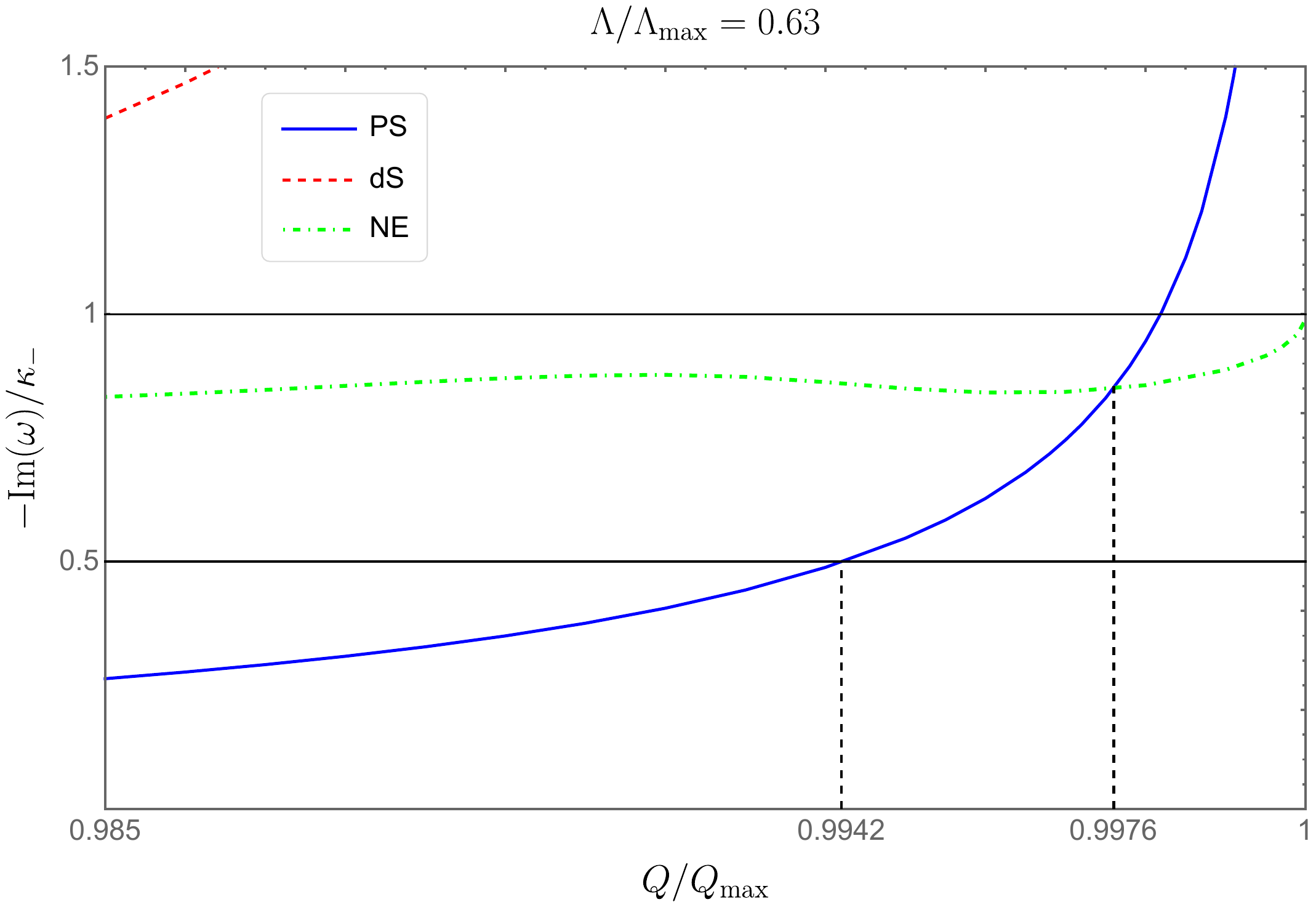}
\includegraphics[height=2.4in,width=3.2in]{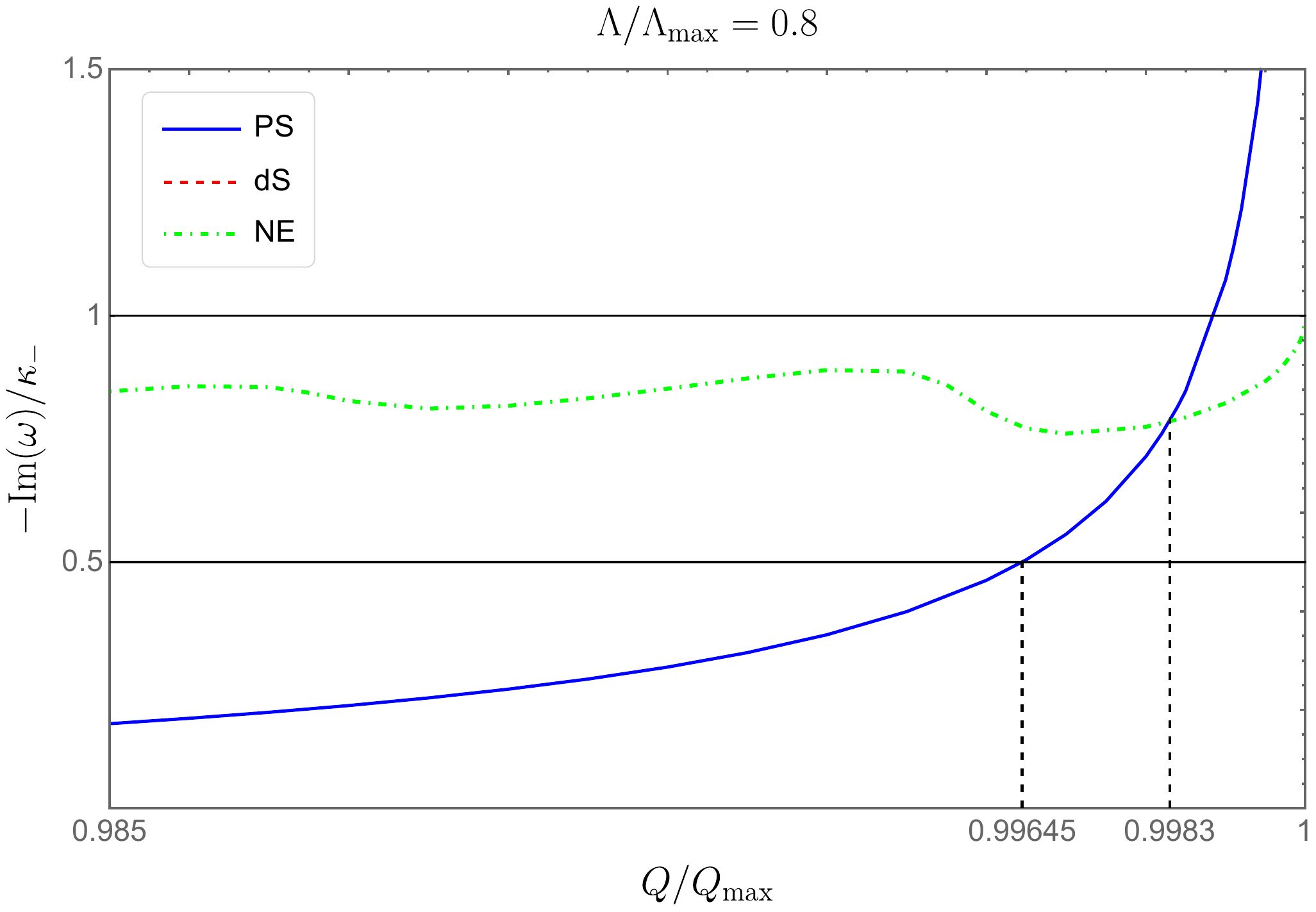}
\caption{The behaviors of $\beta_{\mathrm{PS,dS, NE}}$ related to  fundamental PS modes (blue solid, $l=10$), dS modes (red dashed, $l=1$) and NE modes (green dot-dashed, $l=0$) of massless scalar field perturbation as a function of $Q/Q_{\mathrm{max}}$. In our ABGB-dS black hole case, $\Lambda/\Lambda_{\mathrm{max}}=0.09,0.27,0.63$ correspond to $\Lambda=0.02,0.06,0.14$ in RN-dS spacetime \cite{Cardoso:2017soq}, respectively. The last plot with $\Lambda/\Lambda_{\mathrm{max}}=0.8$ can be compared with the result in \cite{Liu:2019lon}. \label{fig2}}
\end{figure}

As we have discussed above, the validity of SCC is controlled by the value of $\beta$ which is defined by the dominant mode of all non-zero QNMs. Based on the classifications of massless scalar QNMs, the $\beta$ can be found by picking out the dominant one among the fundamental QNMs of PS, dS and NE family. Therefore,  we plot $-\mathrm{Im}\,\omega/\kappa_-$ of fundamental  PS, dS and NE modes (i.e. $\beta_{\mathrm{PS,dS, NE}}$) as a function of $Q/Q_{\mathrm{max}}$ in Fig.~\ref{fig2}, which includes four plots and each plot is assigned a different cosmological constant ratio $\Lambda/\Lambda_{\mathrm{max}}$ matching to the ratio used in \cite{Cardoso:2017soq,Liu:2019lon}, such that we can  directly compare our results  with the previous works.

One can see that each plot shows similar behaviors of $\beta_{\mathrm{PS,dS, NE}}$.  For all plots,  we can always find a charge region in which  $\beta>1/2$  in the near extremal regime, which directly confirms the violation of SCC in ABGB-dS spacetime. As the increase of charge ratio, the $\beta_{\mathrm{PS}}$ and $\beta_{\mathrm{dS}}$ will monotonically grow with the tendency  to infinity and consequently become subdominant, while the $\beta_{\mathrm{NE}}$ is always bounded by $\beta_{\mathrm{NE}}<1$ such that it will be bound to dominate and defines the $\beta$ when the charge ratio is sufficient high. For a larger cosmological constant, a higher charge ratio is required to  violate SCC, and meanwhile the same for NE modes to take over the dominance.  In contrast to $\beta_{\mathrm{PS}}$ and $\beta_{\mathrm{dS}}$ which remain a monotonic relationship with charge ratio regardless of the cosmological constant, the $\beta_{\mathrm{NE}}$ of NE modes is peculiar in the sense that it exhibits a non-monotonic behavior as a function of charge ratio when cosmological constant is not too small, as shown in the lower left and right panels of Fig.~\ref{fig2}. However, this unique property of $\beta_{\mathrm{NE}}$ is ignored in RN-dS spacetime \cite{Cardoso:2017soq,Liu:2019lon,Destounis:2019omd} due to  a relatively narrow range of charge ratio was considered there. On the other hand, we find that $\beta_{\mathrm{dS}}$ of dS modes is the most sensitive  to  cosmological constant, since it  moves upward quickly and so becomes more and more subdominant when increasing the cosmological constant, which just lightly affects $\beta_{\mathrm{PS}}$ by pushing it downward a little bit.

%%%%%%%%%%%%%%%%%%%%%%%%%%%%%%%
%%%%%%%%%%%%%%%%%%%%%%%%%%%%%%%

\begin{figure}[thbp]
\centering
\includegraphics[height=2.8in,width=4.2in]{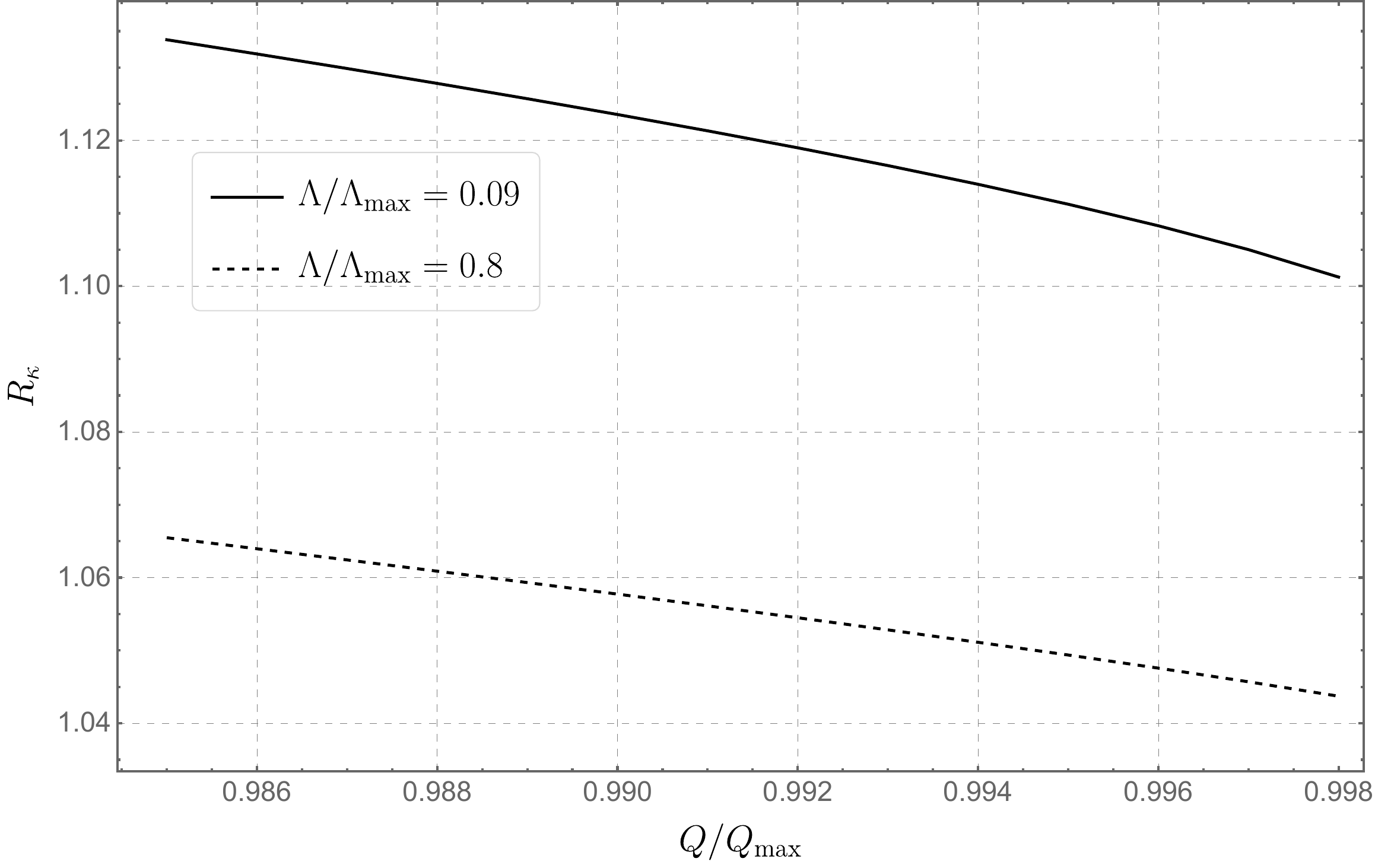}
\caption{The dependence of Cauchy horizon surface gravity ratio $R_k\equiv\kappa^{\rm{RN-dS}}_{-}/\kappa^{\rm{ABGB-dS}}_{-}$  between RN-dS and ABGB-dS black hole on  $Q/Q_{\max}$. \label{fig5}}
\end{figure}

By comparing our results with the results  in RN-dS spacetime \cite{Cardoso:2017soq,Liu:2019lon}, we find it easier to break SCC in this regular ABGB-dS spacetime than that in RN-dS spacetime, in the sense that lower charge ratio is needed to satisfy $\beta>1/2$. For example, in our present case with $\Lambda/\Lambda_{\mathrm{max}}=0.09$, the minimum  charge ratio required to have $\beta>1/2$ is $Q/Q_{\mathrm{max}}=0.9899$, while it would be $Q/Q_{\mathrm{max}}=0.9917$ in RN-dS spacetime (corresponding to $\Lambda=0.02$) as shown in \cite{Cardoso:2017soq}. However, this is not the case for a larger $\Lambda/\Lambda_{\mathrm{max}}=0.8$ under which the situation is reversed. 
The  occurrence of this reversal can be understood by comparing the  surface gravity of Cauchy horizon for RN-dS black hole (denoted by $\kappa^{\rm{RN-dS}}_{-}$) with  surface gravity of Cauchy horizon for ABGB-dS black hole (denoted by $\kappa^{\rm{ABGB-dS}}_{-}$). The comparison is illustrated in Fig.~\ref{fig5} which shows the behaviors of  $R_{\kappa}\equiv\kappa^{\rm{RN-dS}}_{-}/\kappa^{\rm{ABGB-dS}}_{-}$ under the change of $Q/Q_{\max}$ for $\Lambda/\Lambda_{\max}=0.09$ and $0.8$. We see that $R_{\kappa}>1$ holds which means  that  $\kappa^{\rm{RN-dS}}_{-}$ remains bigger than $\kappa^{\rm{ABGB-dS}}_{-}$ in the whole charge range we considered no matter for a smaller or larger cosmological constant. A smaller Cauchy horizon surface gravity for ABGB-dS black hole makes it have an advantage of breaking SCC relative to the RN-dS case, since $\beta=\alpha/\kappa^{\rm{ABGB-dS}}_{-}$ tends to be larger and therefore a smaller $Q/Q_{\max}$ may be required  to achieve $\beta>1/2$. However, $R_\kappa$ approaches unity as $\Lambda/\Lambda_{\max}$ increases from 0.09 to 0.8. This indicates that the relative modification of the Cauchy horizon surface gravity induced by the regularization related to the nonlinear electrodynamics becomes weaker for larger cosmological constant. Consequently, the contribution of the blueshift from the Cauchy horizon to the difference between the two SCC thresholds becomes less pronounced, and the relative modification of the exterior decay rate of QNMs must play an increasingly important role to finally make the reversal happen. In short, when compared with RN-dS black hole at a given large cosmological constant, the effects of nonlinear electrodynamics in ABGB-dS spacetime will make the decay rate of dominant mode  drop faster than the surface gravity of Cauchy horizon such that a higher critical charge ratio is required to satisfy SCC violation criterion $\beta>1/2$. This explains the trend reversal observed at large cosmological constant.

%%%%%%%%%%%%%%%%%%%%%%%%%%%%%%%
%%%%%%%%%%%%%%%%%%%%%%%%%%%%%%%

\subsection{SCC for massive scalar perturbations}

In this part, we would like to investigate the fate of SCC under massive scalar perturbations as a natural extension of massless case. It should be noted that the classifications of massless QNMs introduced previously may not be applicable to the massive scalar QNMs. Hence, in the search of $\beta$,  we take the QNF with $l=0,1,2,3,\dots,10$ as representatives of $\{\omega_{ln}\}$, therefore the $\beta$ is determined by the dominant one among these  QNMs. To further clarify that these QNMs are enough to determine $\beta$, we carry on with our calculations  for QNMs with higher angular  numbers in Appendix~\ref{app1}.

%%%%%%%%%%%%%%%%%%%%%%%%%%%%%%%
\begin{figure}[thbp]
\centering
\includegraphics[height=2.4in,width=3.2in]{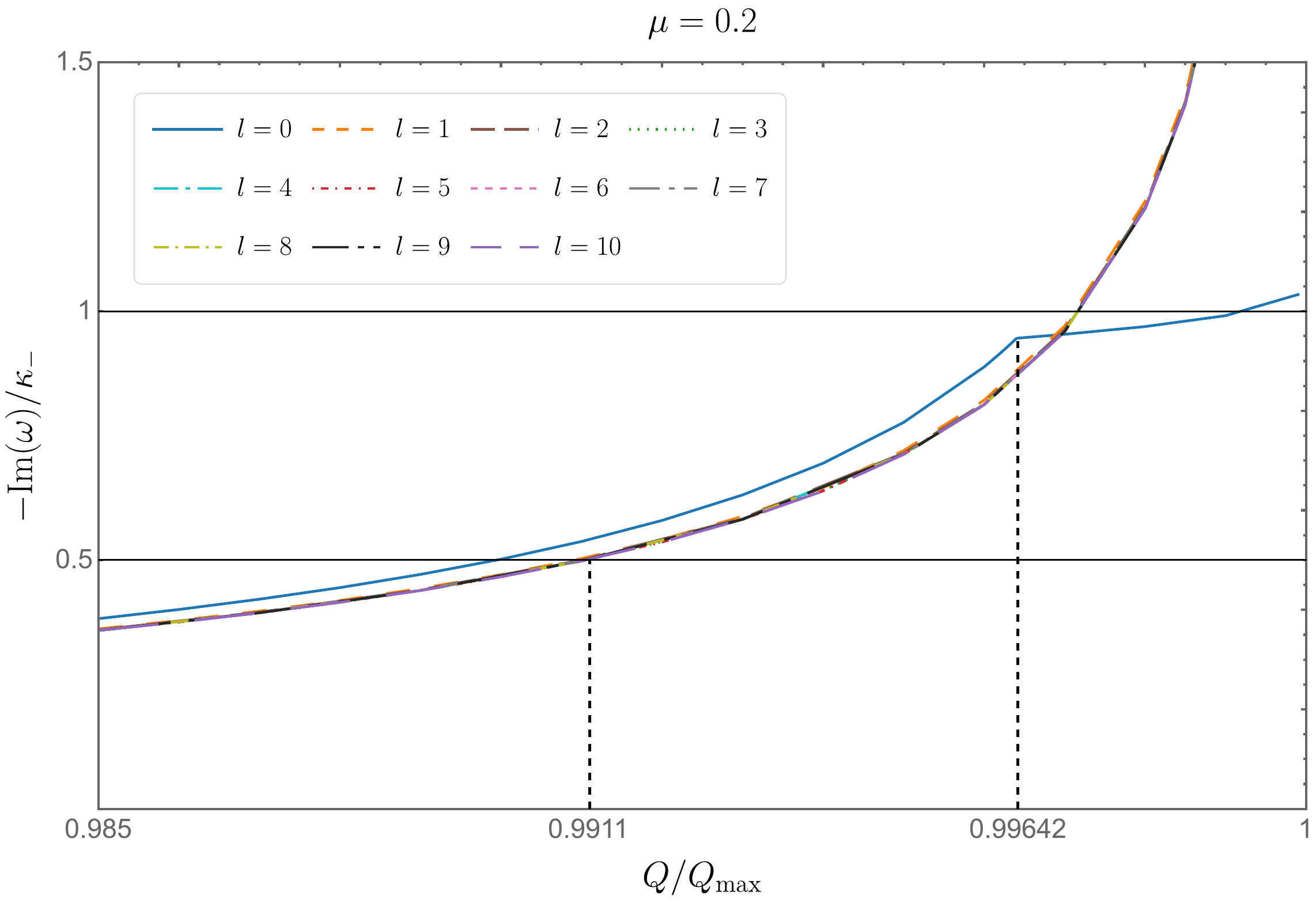}
\includegraphics[height=2.4in,width=3.2in]{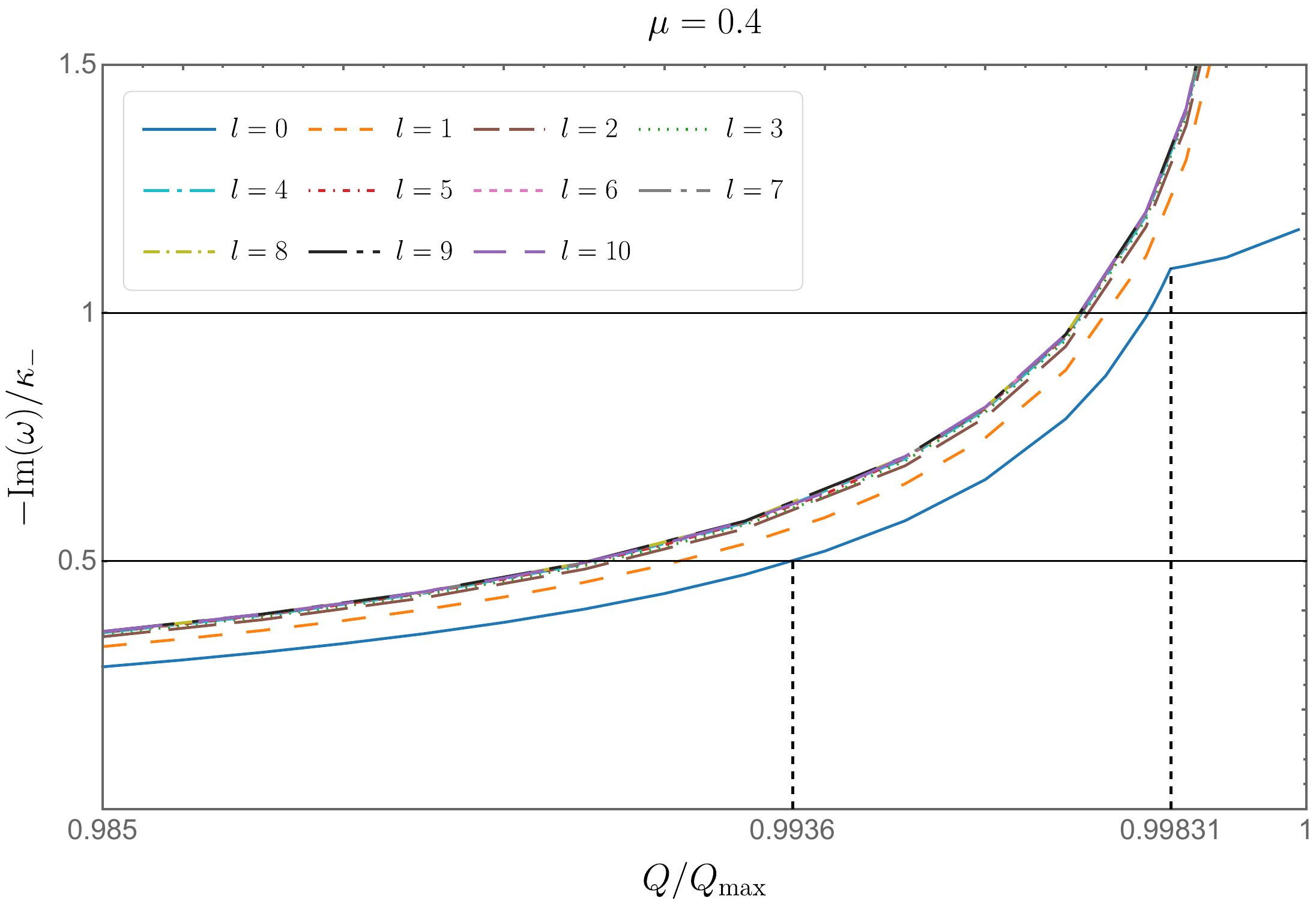}

\includegraphics[height=2.4in,width=3.2in]{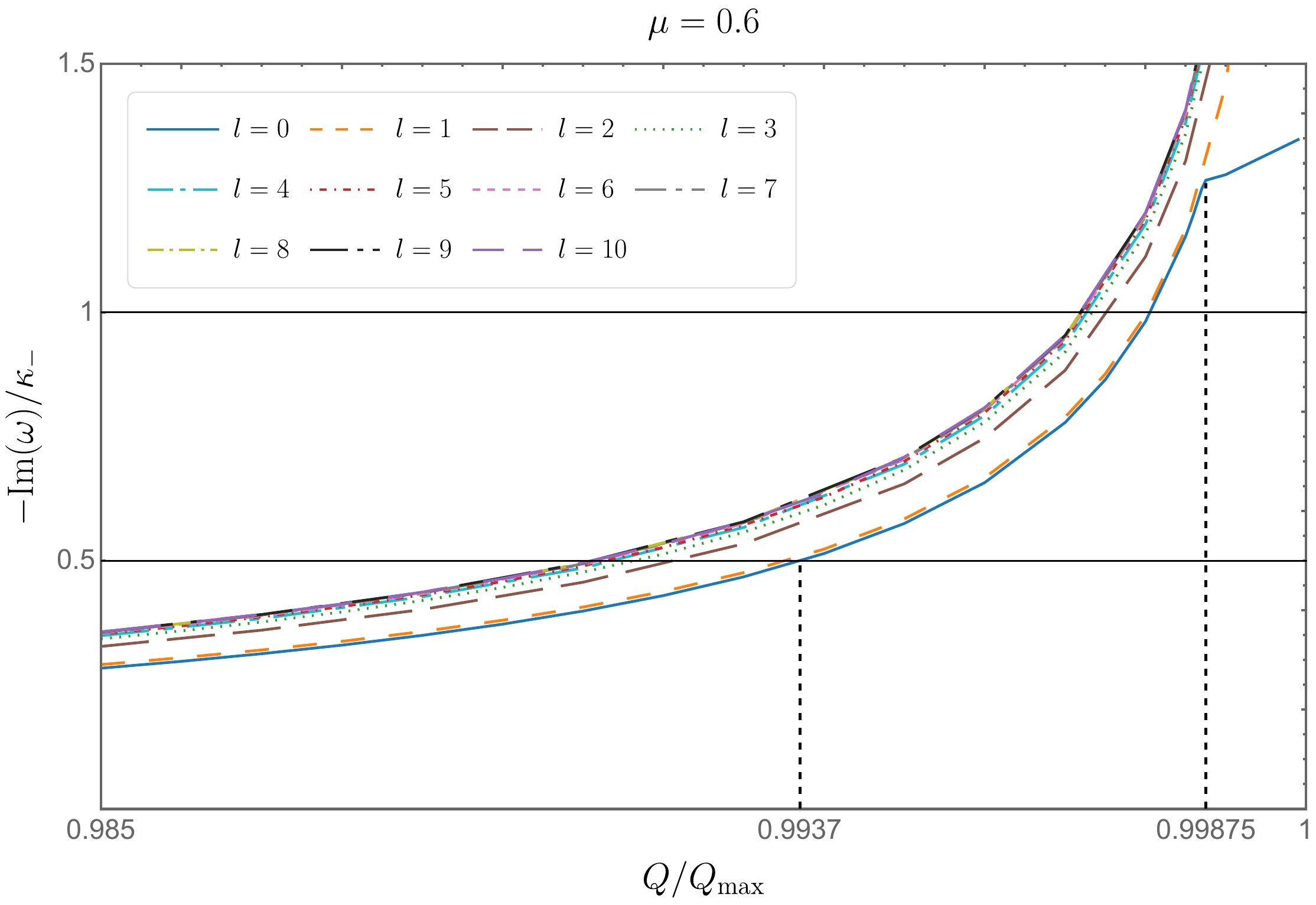}
\includegraphics[height=2.4in,width=3.2in]{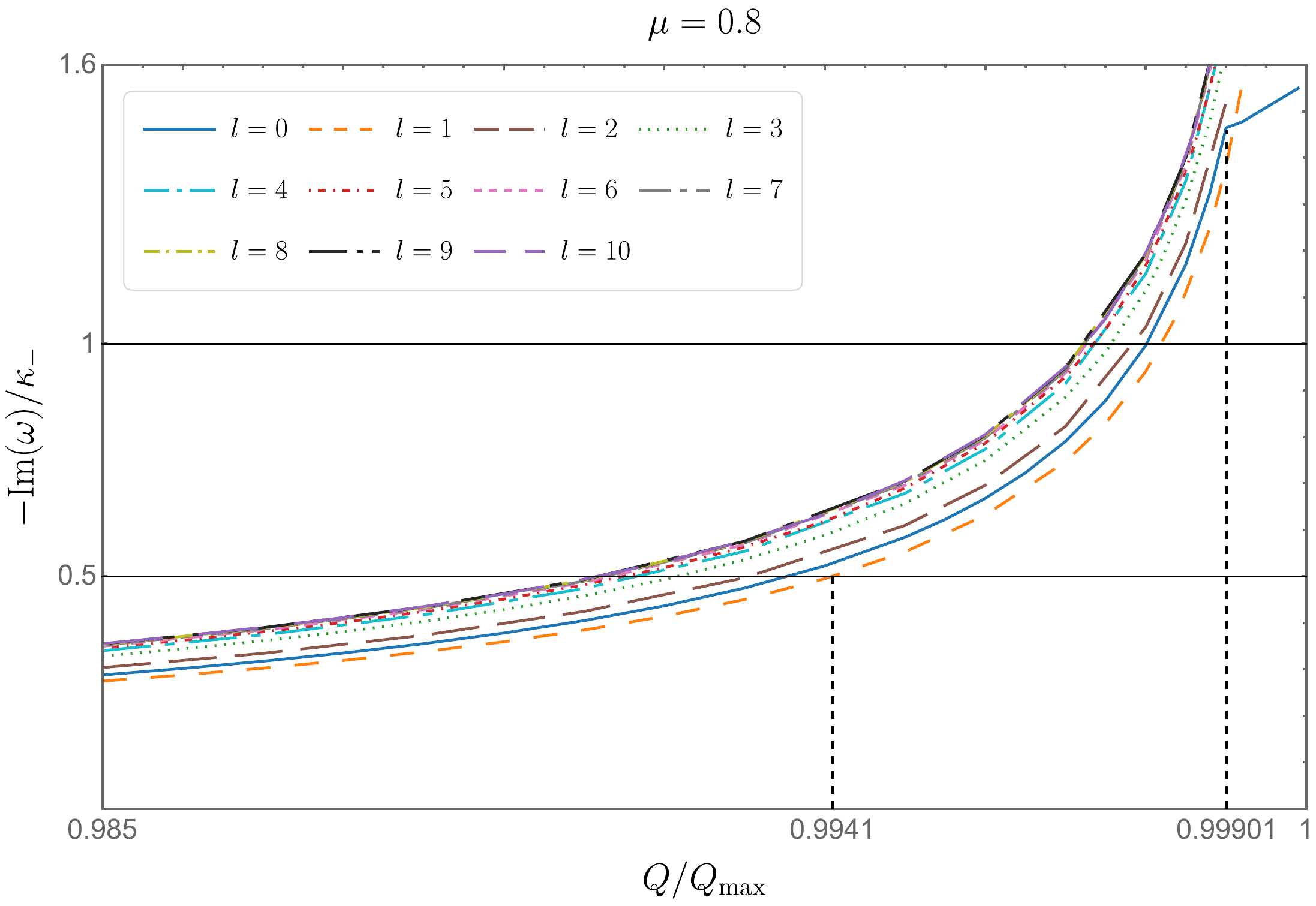}
\caption{The behaviors of $-\mathrm{Im}\,\omega/\kappa_{-}$ from  fundamental  massive QNMs with angular number $l=0,1,2,3,\dots,10$  as a function of $Q/Q_{\mathrm{max}}$ for several scalar mass values at  $\Lambda/\Lambda_{\mathrm{max}}=0.27$.\label{fig3}}
\end{figure}
%%%%%%%%%%%%%%%%%%%%%%%%%%%%%%%

Four plots with different scalar mass are shown in Fig.~\ref{fig3}, which demonstrates the dependence  of $-\mathrm{Im}\,\omega/\kappa_{-}$ of fundamental QNF on the  charge ratio. The figure clearly shows that the scalar mass cannot save SCC, but it indeed increases the difficulty of breaking SCC as higher charge ratio is needed to ensure $\beta>1/2$. For sufficiently high charge ratio, the $l=0$ mode still dominates  and the others  diverge as expected. Nevertheless,  one particular new feature is introduced by the scalar mass. The nondivergence of $-\mathrm{Im}\,\omega/\kappa_{-}$ for $l=0$ modes remains unaffected by scalar mass, but the upper bound is improved, i.e. $\beta>1$ is achieved.  Note that $\beta$ is the measure of the regularity of perturbation field on Cauchy horizon, $\beta>1$ means that the perturbation field can be extended across the Cauchy horizon with $C^{1}$ continuity.

The $C^{1}$ extendibility of the scalar field implies a locally bounded energy-momentum tensor, therefore its coupling to gravity can lead to solutions with bounded Ricci curvature. We should point out that the bounded Ricci curvature alone does not necessarily indicate a bounded Kretschmann scalar, which is  controlled by both the gradient of scalar field and Hawking mass for spherically symmetric self gravitating scalar field \cite{Costa:2014aia}. Whether the Hawking mass is bounded or undergoes mass inflation is a nonlinear question that requires  an analysis of nonlinear backreaction from the coupled gravitation-matter field system. However, in present work we only focus on linear scalar perturbations on a fixed ABGB-dS background without considering any backreaction.  Therefore,  the emergence of  $\beta>1$ may just suggest a substantially weaker curvature singularity and thus of a more severe breakdown of SCC than the case $\beta<1$, rather than the presence of bounded curvature on Cauchy horizon.

\begin{figure}[thbp]
\centering
\includegraphics[height=2.3in,width=3.2in]{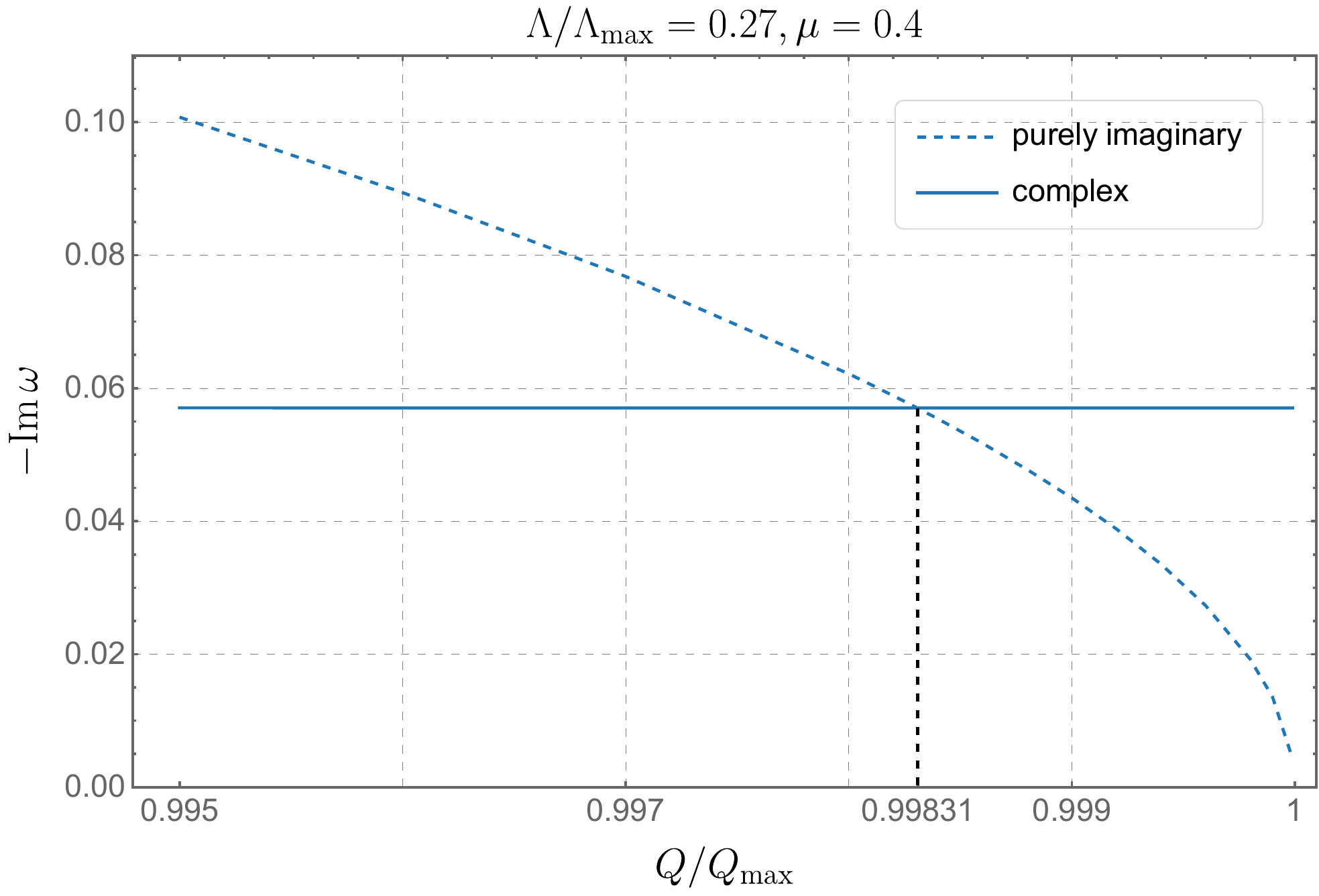}
\includegraphics[height=2.3in,width=3.2in]{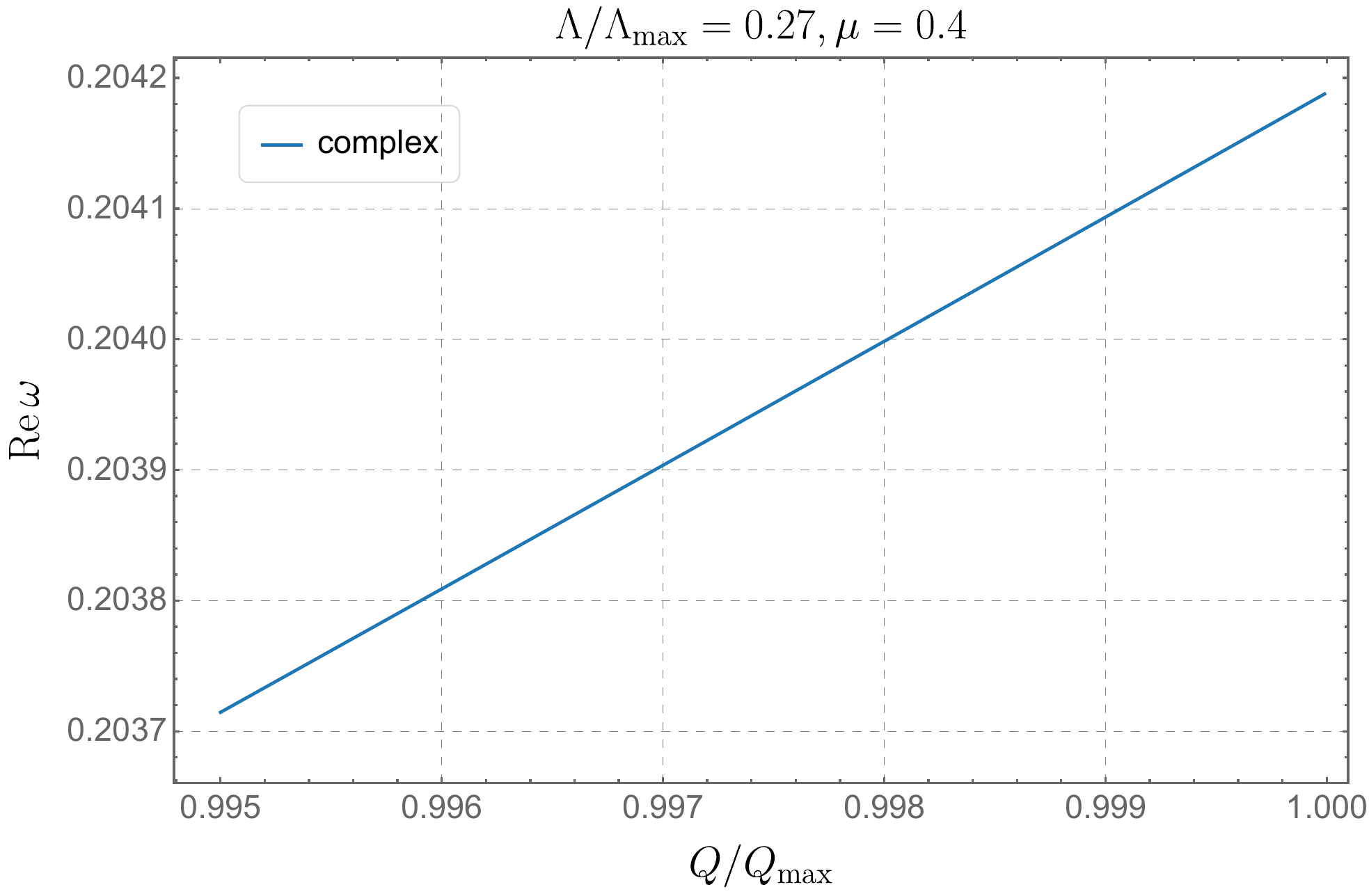}
\caption{The fundamental QNF of  complex and purely imaginary branch of massive QNMs with $l=0$ for $\Lambda/\Lambda_{\mathrm{max}}=0.27, \mu=0.4$. A switching point is located at $Q/Q_{\max}=0.99831$, beyond which the purely imaginary mode will dominate and its QNF quickly converges to $\omega=0$ when approaching extremality and thus guaranteeing a bounded $\beta$.\label{fig6}}
\end{figure}

Furthermore,  non-smoothness is observed for each curve associated with  $l=0$ modes. The occurrence of non-smoothness arises because  the fundamental QNF is dominated by distinct branches of QNMs across  different  regimes of charge ratio (although classification of massive QNMs families is not made). Our numerics indicate that beyond  some critical charge ratio, there are always  purely imaginary  modes showing up (analogous to NE modes in massless case) as dominant ones which render the $\beta$ bounded, and scalar mass can elevate this critical charge ratio. To more explicitly clarify our point, we demonstrate the process of branch switching as the charge ratio changes in Fig.~\ref{fig6}, which  shows the fundamental QNF of complex and purely imaginary mode in high charge ratio region for massive QNMs with $l=0,\Lambda/\Lambda_{\mathrm{max}}=0.27$ and $\mu=0.4$. One can see that the imaginary parts of the  two branches cross at point $Q/Q_{\max}\simeq 0.99831$. The dominant mode will switch  once the charge ratio crosses this point, and this fact accounts for the appearance of non-smoothness observed in Fig.~\ref{fig3} since which just only  concerns the dominant mode under each angular number. This figure also shows us that a purely imaginary branch of QNMs  will arise and serve as the dominant mode when charge ratio is sufficient high, just as the NE mode in massless QNMs case. 

%%%%%%%%%%%%%%%%%%%%%%%%%%%%%%%
%%%%%%%%%%%%%%%%%%%%%%%%%%%%%%%

\section{Conclusions and Discussions}\label{sec5}

In this paper, we have studied the properties of  QNMs of scalar perturbations and SCC in regular ABGB-dS black hole spacetime. The two research topics are connected as the fate of SCC depends on the  spectral gap determined by the late time behaviors of QNMs. 
We considered the influences of scalar mass $\mu$, black hole charge $Q$, cosmological constant $\Lambda$ and angular number $l$ on QNF. We find that $\omega_R$ and $\omega_I$ increase with the scalar mass suggesting that the heavier QNMs oscillate more rapidly but have slower decay rate. The effects of scalar mass are also reflected in the variation of angular number. When increasing angular number, $\omega_R$ becomes larger, while the changes of $\omega_I$ are dependent on the scalar mass value. For a small scalar mass, $\omega_I$  grows with the angular number. If the scalar mass is not that small, $\omega_I$ decreases with the angular number. Especially, we spotted a purely imaginary QNF for $\mu=0.1, l=0$ in Table~\ref{tab1}. We find that this mode is traced back to the zero-mode in pure dS spacetime and it is the scalar mass that makes it deviate from  $\omega=0$. Actually, this mode does not remain purely imaginary, instead it will become complex when the scalar mass is large enough, and it behaves similarly to massive QNMs with $l=0,n=0$ in pure dS spacetime, therefore we call it dS-like mode. For larger scalar mass ($\mu\gtrsim0.1812$), dS-like mode becomes subdominant thus the  complex QNF serves as fundamental QNF, as shown in Table~\ref{tab1} and Fig.~\ref{fig1}. 
When we increase black hole charge $Q$, it is found that $\omega_R$ grows but $\omega_I$ decreases, hence QNMs will oscillate more rapidly and decay faster for a larger value of Q. For cosmological constant $\Lambda$, on the contrary to charge, we find that a larger $\Lambda$ will reduce $\omega_R$ and in the meantime increase the $\omega_I$ leading to a lower oscillation frequency and damping rate. In this sense, the cosmological constant and charge modify the QNF in opposite way. When considering the angular number, we note that it changes the $\omega_I$ in different manner depending on the value of $\Lambda$. For a small $\Lambda=0.05$, larger angular number lowers  $\omega_I$ and thus the  $l=0$ mode is the dominant one. However,  this is not the case for bigger $\Lambda$ under which the dominant mode corresponds to large $l$.

We have investigated SCC separately for massless and massive scalar perturbations, and found that the SCC  can be violated in near-extremal  regular ABGB-dS black hole spacetime no matter the scalar perturbations are massless or massive. For the massless perturbations, we have shown that the QNMs can be classified into three families, namely PS, dS and NE modes and they behave differently from each other as the charge ratio varies. Both $\beta_{\mathrm{PS}}$ and $\beta_{\mathrm{dS}}$ will go to infinity as the black hole approaches extremality ($Q/Q_{\mathrm{max}}\to1$), whilst $\beta_{\mathrm{NE}}$ is bounded by $\beta_{\mathrm{NE}}<1$ suggesting that the NE modes always dominate and define the $\beta$ at sufficiently high charge ratio and the curvature on Cauchy horizon blows up (note that $\beta$ measures the regularity of perturbations). On the other hand, the NE modes exhibit a nonmonotonic behavior as a function of charge ratio when cosmological constant is not too small. This unique feature  of NE modes seems to be overlooked in  RN-dS case \cite{Cardoso:2017soq,Liu:2019lon,Destounis:2019omd}.  When compared with RN-dS spacetime \cite{Cardoso:2017soq,Liu:2019lon}, we find it easier to break SCC in this regular ABGB-dS spacetime, in the sense that lower charge ratio is needed to satisfy $\beta>1/2$. However, this is not the case for a larger $\Lambda/\Lambda_{\mathrm{max}}=0.8$, by which the situation is reversed. 

For massive scalar perturbations, we find that the SCC cannot be restored by scalar mass, but it indeed enhances the difficulty of breaking SCC as higher charge ratio is needed to ensure $\beta>1/2$. Actually, scalar mass leads to a more severe violation of SCC in the sense that $\beta>1$ is reached which indicates a $C^1$ continuity of perturbation field across the Cauchy horizon, which then may have a weaker curvature singularity and an improved geometric regularity. We have observed non-smoothness of the curves of $-\mathrm{Im}\,\omega/\kappa_-$ for massive $l=0$ modes. This non-smoothness behavior is related to the fact that the fundamental QNF originates from different types of QNMs, and it is  the charge ratio that determines which type of QNMs dominate. Accordingly, there exists some  critical charge ratio which divide the region where different QNMs dominate. Beyond this critical  charge ratio, we find that the purely imaginary  modes (similar to NE modes in massless case) always emerge and take over the dynamics as dominant modes, and this critical charge ratio can be elevated by scalar mass.

The SCC violation identified in this work provides a concrete case for the considerations raised in the Introduction. SCC violation in a regular black hole is not a purely formal issue, it calls for an analysis of the physics beyond the Cauchy horizon. In particular, regular black holes, unprotected by weak cosmic censorship, can lose their event horizons~\cite{Li:2013sea,Yang:2022yvq} and expose their interiors to observers, yielding potentially detectable electromagnetic and gravitational wave signals~\cite{Cao:2023par,Fauzi:2025ldu}, or providing us the possibility to access the quantum regime of gravity inside black holes \cite{Yang:2022yvq} if the SCC can be violated. Whereas, if the SCC is always respected in the realistic gravitational collapse leading to a black hole, then  the spacetime manifold may end at Cauchy horizon such that the regular core of the static ABGB-dS spacetime may fail to be dynamically  realized, let alone exploring the physics responsible for the singularity resolution. These prospects bear directly on our understanding of singularities and quantum gravity~\cite{Ashtekar:2005qt,Bojowald:2018xxu,Ashtekar:2018lag}. 

However, we should keep in mind  that  our analysis is limited to the linear level. A value $\beta>1/2$ just indicates that the linear perturbations satisfy the regularity condition associated  with a possible weak solution of Einstein's equation  across the Cauchy horizon, but it does not by itself establish the nonlinear stability of the horizon or the existence of a regular extension of the spacetime which actually undergoes fully nonlinear dynamical evolution. If such extendibility survives when gravitational backreaction is included, regular black holes would indeed provide an especially interesting setting because the extension could in principle connect to a nonsingular interior and the physics therein  rather than to a central curvature singularity, as the outlooks  we have discussed above. Whether the regular core is physically accessible, therefore  remains an open question requiring nonlinear gravitational analysis.

\begin{acknowledgments}
This work is supported by  National Natural Science Foundation of China under Grant No.12305071.
\end{acknowledgments}

\appendix

\section{Convergence of the massive QNMs with high angular numbers}\label{app1}

To make sure that we have not missed any QNMs which are much more dominant than the modes illustrated  in Fig.~\ref{fig3} and therefore  the spectral gap may be significantly changed, we calculate $-\mathrm{Im}\,\omega/\kappa_{-}$ of QNMs with angular number $l=10,11,\dots,20$ and $l\to\infty$ for scalar mass $\mu=0.2,0.4,0.6$ and $\mu=0.8$ at $\Lambda/\Lambda_{\max}=0.27$.  The numerical results obtained by PSM (for finite $l$) and eikonal limit formula Eq.~\eqref{ps formula} (for $l\to\infty$) are listed in Table~\ref{tabapp1}, Table~\ref{tabapp2}, Table~\ref{tabapp3}, and Table~\ref{tabapp4}.  The numerics  show  that the  QNMs with $l\geq10$ can be used to represent QNMs in eikonal limit  since all these modes are  well approximated by $l\to\infty$ limit. We therefore believe that the modes we used in Fig.~\ref{fig3} are sufficient to determine $\beta$.

%%%%%%%%%%%%%%%%%%%%%%%%%%%%%%%
\begin{table}[!htbp]
\centering
\setlength{\tabcolsep}{24pt}
\caption*{$\mu=0.2$} 
%\resizebox{\textwidth}{!}
{
        \begin{tabular}{ccccc}
    \hline    % after \\: \hline or \cline{col1-col2} \cline{col3-col4} ...
    \hline
    $l$  &         $q=0.985$         & $q=0.998$           & $q=0.999$              &  $q=0.9999$      \\
      
         \hline
                        
     10  &         $0.358630 $       & $1.20601 $          & $1.76652 $             & $5.92679 $  \\
                     
     11  &         $0.358619 $       & $1.20595 $          & $1.76644 $             & $5.92652 $  \\
                       
     12  &         $0.358611 $       & $1.20592 $          & $1.76638 $             & $5.92631 $  \\
                        
     13  &         $0.358605 $       & $1.20588 $          & $1.76633 $             & $5.92614 $  \\
                        
     14  &         $0.358599 $       & $1.20586 $          & $1.76629 $             & $5.92601 $  \\
                       
     15  &         $0.358595 $       & $1.20584 $          & $1.76626 $             & $5.92590 $  \\
          
     16  &         $0.358592 $       & $1.20582 $          & $1.76623 $             & $5.92581 $  \\
                       
     17  &         $0.358589 $       & $1.20581 $          & $1.76621 $             & $5.92573 $  \\
                       
     18  &         $0.358587 $       & $1.20580 $          & $1.76620 $             & $5.92567 $  \\
                        
     19  &         $0.358584 $       & $1.20579 $          & $1.76618 $             & $5.92562 $  \\
                        
     20  &         $0.358583 $       & $1.20578 $          & $1.76617 $             & $5.92557 $  \\
     
  $l\to\infty$ &   $0.358566 $       & $1.20570 $          & $1.76604 $             & $5.92514 $  \\

           \hline
           \hline
           \end{tabular}
}
\caption{The values of $-\mathrm{Im}\,\omega/\kappa_{-}$ of fundamental  massive QNMs with angular number $l=10,11,\dots,20$  and $l\to \infty$  for several charge ratios $q\equiv Q/Q_{\max}$ at  $\Lambda/\Lambda_{\mathrm{max}}=0.27,\mu=0.2$. }\label{tabapp1}
\end{table}
%%%%%%%%%%%%%%%%%%%%%%%%%%%%%%%

%%%%%%%%%%%%%%%%%%%%%%%%%%%%%%%
\begin{table}[!htbp]
\centering
\setlength{\tabcolsep}{24pt}
\caption*{$\mu=0.4$} 
%\resizebox{\textwidth}{!}
{
        \begin{tabular}{ccccc}
    \hline    % after \\: \hline or \cline{col1-col2} \cline{col3-col4} ...
    \hline
    $l$  &         $q=0.985$         & $q=0.998$           & $q=0.999$              &  $q=0.9999$      \\
      
         \hline
                        
     10  &         $0.357934 $       & $1.20396 $          & $1.76356 $             & $5.91698 $  \\
                     
     11  &         $0.358039 $       & $1.20425 $          & $1.76397 $             & $5.91834 $  \\
                       
     12  &         $0.358120 $       & $1.20447 $          & $1.76429 $             & $5.91939 $  \\
                        
     13  &         $0.358184 $       & $1.20465 $          & $1.76454 $             & $5.92021 $  \\
                        
     14  &         $0.358235 $       & $1.20479 $          & $1.76474 $             & $5.92087 $  \\
                       
     15  &         $0.358276 $       & $1.20490 $          & $1.76490 $             & $5.92140 $  \\
          
     16  &         $0.358310 $       & $1.20499 $          & $1.76504 $             & $5.92184 $  \\
                       
     17  &         $0.358339 $       & $1.20507 $          & $1.76515 $             & $5.92221 $  \\
                       
     18  &         $0.358362 $       & $1.20514 $          & $1.76524 $             & $5.92252 $  \\
                        
     19  &         $0.358383 $       & $1.20519 $          & $1.76532 $             & $5.92278 $  \\
                        
     20  &         $0.358400 $       & $1.20524 $          & $1.76539 $             & $5.92300 $  \\
     
  $l\to\infty$ &   $0.358566 $       & $1.20570 $          & $1.76604 $             & $5.92514 $  \\

           \hline
           \hline
           \end{tabular}
}
\caption{The values of $-\mathrm{Im}\,\omega/\kappa_{-}$ of fundamental  massive QNMs with angular number $l=10,11,\dots,20$ and $l\to \infty$  for several charge ratios $q\equiv Q/Q_{\max}$ at  $\Lambda/\Lambda_{\mathrm{max}}=0.27,\mu=0.4$.  }\label{tabapp2}
\end{table}
%%%%%%%%%%%%%%%%%%%%%%%%%%%%%%%

%%%%%%%%%%%%%%%%%%%%%%%%%%%%%%%
\begin{table}[!htbp]
\centering
\setlength{\tabcolsep}{24pt}
\caption*{$\mu=0.6$} 
%\resizebox{\textwidth}{!}
{
        \begin{tabular}{ccccc}
    \hline    % after \\: \hline or \cline{col1-col2} \cline{col3-col4} ...
    \hline
    $l$  &         $q=0.985$         & $q=0.998$           & $q=0.999$              & $q=0.9999$      \\
      
         \hline
                        
     10  &         $0.356775 $       & $1.20054 $          & $1.75861 $             & $5.90061 $  \\
                     
     11  &         $0.357073 $       & $1.20140 $          & $1.75985 $             & $5.90470 $  \\
                       
     12  &         $0.357302 $       & $1.20206 $          & $1.76080 $             & $5.90785 $  \\
                        
     13  &         $0.357482 $       & $1.20258 $          & $1.76155 $             & $5.91032 $  \\
                        
     14  &         $0.357627 $       & $1.20299 $          & $1.76215 $             & $5.91230 $  \\
                       
     15  &         $0.357744 $       & $1.20333 $          & $1.76264 $             & $5.91390 $  \\
          
     16  &         $0.357841 $       & $1.20361 $          & $1.76304 $             & $5.91522 $  \\
                       
     17  &         $0.357921 $       & $1.20384 $          & $1.76337 $             & $5.91633 $  \\
                       
     18  &         $0.357989 $       & $1.20404 $          & $1.76365 $             & $5.91725 $  \\
                        
     19  &         $0.358047 $       & $1.20420 $          & $1.76389 $             & $5.91804 $  \\
                        
     20  &         $0.358096 $       & $1.20435 $          & $1.76410 $             & $5.91872 $  \\
     
  $l\to\infty$ &   $0.358566 $       & $1.20570 $          & $1.76604 $             & $5.92514 $  \\

           \hline
           \hline
           \end{tabular}
}
\caption{The values of $-\mathrm{Im}\,\omega/\kappa_{-}$ of fundamental  massive QNMs with angular number $l=10,11,\dots,20$  and $l\to \infty$  for several  charge ratios $q\equiv Q/Q_{\max}$ at  $\Lambda/\Lambda_{\mathrm{max}}=0.27,\mu=0.6$.  }\label{tabapp3}
\end{table}
%%%%%%%%%%%%%%%%%%%%%%%%%%%%%%%

%%%%%%%%%%%%%%%%%%%%%%%%%%%%%%%
\begin{table}[!htbp]
\centering
\setlength{\tabcolsep}{24pt}
\caption*{$\mu=0.8$} 
%\resizebox{\textwidth}{!}
{
        \begin{tabular}{ccccc}
    \hline    % after \\: \hline or \cline{col1-col2} \cline{col3-col4} ...
    \hline
    $l$  &         $q=0.985$         & $q=0.998$           & $q=0.999$              &  $q=0.9999$      \\
      
         \hline
                        
     10  &         $0.355152 $       & $1.19574 $          & $1.75167 $             & $5.87762 $  \\
                     
     11  &         $0.355720 $       & $1.19740 $          & $1.75407 $             & $5.88556 $  \\
                       
     12  &         $0.356157 $       & $1.19868 $          & $1.75592 $             & $5.89166 $  \\
                        
     13  &         $0.356501 $       & $1.19968 $          & $1.75737 $             & $5.89645 $  \\
                        
     14  &         $0.356776 $       & $1.20048 $          & $1.75852 $             & $5.90028 $  \\
                       
     15  &         $0.356999 $       & $1.20114 $          & $1.75947 $             & $5.90339 $  \\
          
     16  &         $0.357184 $       & $1.20167 $          & $1.76024 $             & $5.90595 $  \\
                       
     17  &         $0.357337 $       & $1.20212 $          & $1.76089 $             & $5.90809 $  \\
                       
     18  &         $0.357466 $       & $1.20250 $          & $1.76143 $             & $5.90988 $  \\
                        
     19  &         $0.357576 $       & $1.20282 $          & $1.76189 $             & $5.91141 $  \\
                        
     20  &         $0.357670 $       & $1.20309 $          & $1.76229 $             & $5.91272 $  \\
     
  $l\to\infty$ &   $0.358566 $       & $1.20570 $          & $1.76604 $             & $5.92514 $  \\

           \hline
           \hline
           \end{tabular}
}
\caption{The values of $-\mathrm{Im}\,\omega/\kappa_{-}$ of fundamental  massive QNMs with angular number $l=10,11,\dots,20$  and $l\to \infty$  for several charge ratios $q\equiv Q/Q_{\max}$ at  $\Lambda/\Lambda_{\mathrm{max}}=0.27,\mu=0.8$.  }\label{tabapp4}
\end{table}
%%%%%%%%%%%%%%%%%%%%%%%%%%%%%%%

\bibliographystyle{JHEP}
%\bibliographystyle{apsrev4-1}
%注意tex文件名不能有空格否则参考文献识别不出来！！！！！！！！！！！！！！！！
%\clearpage
\bibliography{References_SCC}

@Article{Matyjasek:2008kn,
  author        = {Matyjasek, Jerzy and Tryniecki, Dariusz and Klimek, Mariusz},
  journal       = {Mod. Phys. Lett. A},
  title         = {{Regular black holes in an asymptotically de Sitter universe}},
  year          = {2009},
  pages         = {3377--3392},
  volume        = {23},
  archiveprefix = {arXiv},
  doi           = {10.1142/S0217732308028715},
  eprint        = {0809.2275},
  primaryclass  = {gr-qc},
}

@Article{Cho:2011sf,
  author        = {Cho, H. T. and Cornell, A. S. and Doukas, Jason and Huang, T. R. and Naylor, Wade},
  journal       = {Adv. Math. Phys.},
  title         = {{A New Approach to Black Hole Quasinormal Modes: A Review of the Asymptotic Iteration Method}},
  year          = {2012},
  pages         = {281705},
  volume        = {2012},
  archiveprefix = {arXiv},
  doi           = {10.1155/2012/281705},
  eprint        = {1111.5024},
  primaryclass  = {gr-qc},
  reportnumber  = {YITP-11-97, WITS-CTP-83, OU-HET-735-2011},
}

@Article{Jansen:2017oag,
  author        = {Jansen, Aron},
  journal       = {Eur. Phys. J. Plus},
  title         = {{Overdamped modes in Schwarzschild-de Sitter and a Mathematica package for the numerical computation of quasinormal modes}},
  year          = {2017},
  number        = {12},
  pages         = {546},
  volume        = {132},
  archiveprefix = {arXiv},
  doi           = {10.1140/epjp/i2017-11825-9},
  eprint        = {1709.09178},
  primaryclass  = {gr-qc},
}

@Article{Konoplya:2019hlu,
  author        = {Konoplya, R. A. and Zhidenko, A. and Zinhailo, A. F.},
  journal       = {Class. Quant. Grav.},
  title         = {{Higher order WKB formula for quasinormal modes and grey-body factors: recipes for quick and accurate calculations}},
  year          = {2019},
  pages         = {155002},
  volume        = {36},
  archiveprefix = {arXiv},
  doi           = {10.1088/1361-6382/ab2e25},
  eprint        = {1904.10333},
  primaryclass  = {gr-qc},
}

@Article{Matyjasek:2017psv,
  author        = {Matyjasek, Jerzy and Opala, Micha{\l}},
  journal       = {Phys. Rev. D},
  title         = {{Quasinormal modes of black holes. The improved semianalytic approach}},
  year          = {2017},
  number        = {2},
  pages         = {024011},
  volume        = {96},
  archiveprefix = {arXiv},
  doi           = {10.1103/PhysRevD.96.024011},
  eprint        = {1704.00361},
  primaryclass  = {gr-qc},
}

@Article{Du:2004jt,
  author        = {Du, Da-Ping and Wang, Bin and Su, Ru-Keng},
  journal       = {Phys. Rev. D},
  title         = {{Quasinormal modes in pure de Sitter space-times}},
  year          = {2004},
  pages         = {064024},
  volume        = {70},
  archiveprefix = {arXiv},
  doi           = {10.1103/PhysRevD.70.064024},
  eprint        = {hep-th/0404047},
}

@Article{Lopez-Ortega:2006aal,
  author        = {Lopez-Ortega, A.},
  journal       = {Gen. Rel. Grav.},
  title         = {{Quasinormal modes of D-dimensional de Sitter spacetime}},
  year          = {2006},
  pages         = {1565--1591},
  volume        = {38},
  archiveprefix = {arXiv},
  doi           = {10.1007/s10714-006-0335-9},
  eprint        = {gr-qc/0605027},
}

@Article{Cardoso:2017soq,
  author        = {Cardoso, Vitor and Costa, Jo{\~a}o L. and Destounis, Kyriakos and Hintz, Peter and Jansen, Aron},
  journal       = {Phys. Rev. Lett.},
  title         = {{Quasinormal modes and Strong Cosmic Censorship}},
  year          = {2018},
  number        = {3},
  pages         = {031103},
  volume        = {120},
  archiveprefix = {arXiv},
  doi           = {10.1103/PhysRevLett.120.031103},
  eprint        = {1711.10502},
  primaryclass  = {gr-qc},
}

@Article{Cardoso:2008bp,
  author        = {Cardoso, Vitor and Miranda, Alex S. and Berti, Emanuele and Witek, Helvi and Zanchin, Vilson T.},
  journal       = {Phys. Rev. D},
  title         = {{Geodesic stability, Lyapunov exponents and quasinormal modes}},
  year          = {2009},
  number        = {6},
  pages         = {064016},
  volume        = {79},
  archiveprefix = {arXiv},
  doi           = {10.1103/PhysRevD.79.064016},
  eprint        = {0812.1806},
  primaryclass  = {hep-th},
}

@Article{Vasy:2007tda,
  author        = {Vasy, Andras},
  journal       = {Adv. Math.},
  title         = {{The wave equation on asymptotically de Sitter-like spaces}},
  year          = {2010},
  pages         = {49--97},
  volume        = {223},
  archiveprefix = {arXiv},
  doi           = {10.1016/j.aim.2009.07.005},
  eprint        = {0706.3669},
  primaryclass  = {math.AP},
}

@Article{Liu:2019lon,
  author        = {Liu, Hang and Tang, Ziyu and Destounis, Kyriakos and Wang, Bin and Papantonopoulos, Eleftherios and Zhang, Hongbao},
  journal       = {JHEP},
  title         = {{Strong Cosmic Censorship in higher-dimensional Reissner-Nordstr{\"o}m-de Sitter spacetime}},
  year          = {2019},
  pages         = {187},
  volume        = {03},
  archiveprefix = {arXiv},
  doi           = {10.1007/JHEP03(2019)187},
  eprint        = {1902.01865},
  primaryclass  = {gr-qc},
}

@Article{Destounis:2019omd,
  author        = {Destounis, Kyriakos and Fontana, Rodrigo D. B. and Mena, Filipe C. and Papantonopoulos, Eleftherios},
  journal       = {JHEP},
  title         = {{Strong Cosmic Censorship in Horndeski Theory}},
  year          = {2019},
  pages         = {280},
  volume        = {10},
  archiveprefix = {arXiv},
  doi           = {10.1007/JHEP10(2019)280},
  eprint        = {1908.09842},
  primaryclass  = {gr-qc},
}

@Article{Dafermos:2003wr,
  author        = {Dafermos, Mihalis},
  journal       = {Commun. Pure Appl. Math.},
  title         = {{The interior of charged black holes and the problem of uniqueness in general relativity}},
  year          = {2005},
  number        = {4},
  pages         = {0445--0504},
  volume        = {58},
  archiveprefix = {arXiv},
  doi           = {10.1002/cpa.20071},
  eprint        = {gr-qc/0307013},
}

@Article{Ori:1991zz,
  author  = {Ori, Amos},
  journal = {Phys. Rev. Lett.},
  title   = {{Inner structure of a charged black hole: An exact mass-inflation solution}},
  year    = {1991},
  pages   = {789--792},
  volume  = {67},
  doi     = {10.1103/PhysRevLett.67.789},
}

@Article{Dafermos:2017dbw,
  author        = {Dafermos, Mihalis and Luk, Jonathan},
  journal       = {Ann. Math. (2)},
  title         = {{The interior of dynamical vacuum black holes. I: The C0-stability of the Kerr Cauchy horizon}},
  year          = {2025},
  number        = {2},
  pages         = {309--630},
  volume        = {202},
  archiveprefix = {arXiv},
  doi           = {10.4007/annals.2025.202.2.1},
  eprint        = {1710.01722},
  primaryclass  = {gr-qc},
}

@Article{VandeMoortel:2025ngd,
  author        = {Van de Moortel, Maxime},
  title         = {{The Strong Cosmic Censorship Conjecture}},
  year          = {2025},
  month         = {1},
  archiveprefix = {arXiv},
  eprint        = {2501.13180},
  primaryclass  = {gr-qc},
}

@InProceedings{Christodoulou:2008nj,
  author        = {Christodoulou, Demetrios},
  booktitle     = {{12th Marcel Grossmann Meeting on General Relativity}},
  title         = {{The Formation of Black Holes in General Relativity}},
  year          = {2008},
  month         = {5},
  pages         = {24--34},
  archiveprefix = {arXiv},
  doi           = {10.1142/9789814374552_0002},
  eprint        = {0805.3880},
  primaryclass  = {gr-qc},
}

@InBook{Penrose:1980ge,
  author    = {Penrose, R.},
  pages     = {581--638},
  title     = {{SINGULARITIES AND TIME ASYMMETRY}},
  year      = {1980},
  booktitle = {{General Relativity}: {An Einstein Centenary Survey}},
}

@InProceedings{1974IAUS...64...82P,
  author    = {{Penrose}, R.},
  booktitle = {Gravitational Radiation and Gravitational Collapse},
  title     = {{Gravitational Collapse (invited Paper)}},
  year      = {1974},
  editor    = {{Dewitt-Morette}, C.},
  month     = jan,
  pages     = {82},
  series    = {IAU Symposium},
  volume    = {64},
  adsurl    = {https://ui.adsabs.harvard.edu/abs/1974IAUS...64...82P},
}

@Article{Luk:2015qja,
  author        = {Luk, Jonathan and Oh, Sung-Jin},
  journal       = {Duke Math. J.},
  title         = {{Proof of linear instability of the Reissner{\textendash}Nordstr{\"o}m Cauchy horizon under scalar perturbations}},
  year          = {2017},
  number        = {3},
  pages         = {437--493},
  volume        = {166},
  archiveprefix = {arXiv},
  doi           = {10.1215/00127094-3715189},
  eprint        = {1501.04598},
  primaryclass  = {gr-qc},
}

@Article{Dafermos:2015bzz,
  author        = {Dafermos, Mihalis and Shlapentokh-Rothman, Yakov},
  journal       = {Commun. Math. Phys.},
  title         = {{Time-Translation Invariance of Scattering Maps and Blue-Shift Instabilities on Kerr Black Hole Spacetimes}},
  year          = {2017},
  number        = {3},
  pages         = {985--1016},
  volume        = {350},
  archiveprefix = {arXiv},
  doi           = {10.1007/s00220-016-2771-z},
  eprint        = {1512.08260},
  primaryclass  = {gr-qc},
}

@Article{Dias:2018ufh,
  author        = {Dias, Oscar J. C. and Reall, Harvey S. and Santos, Jorge E.},
  journal       = {Class. Quant. Grav.},
  title         = {{Strong cosmic censorship for charged de Sitter black holes with a charged scalar field}},
  year          = {2019},
  number        = {4},
  pages         = {045005},
  volume        = {36},
  archiveprefix = {arXiv},
  doi           = {10.1088/1361-6382/aafcf2},
  eprint        = {1808.04832},
  primaryclass  = {gr-qc},
}

@Article{Dias:2018ynt,
  author        = {Dias, Oscar J. C. and Eperon, Felicity C. and Reall, Harvey S. and Santos, Jorge E.},
  journal       = {Phys. Rev. D},
  title         = {{Strong cosmic censorship in de Sitter space}},
  year          = {2018},
  number        = {10},
  pages         = {104060},
  volume        = {97},
  archiveprefix = {arXiv},
  doi           = {10.1103/PhysRevD.97.104060},
  eprint        = {1801.09694},
  primaryclass  = {gr-qc},
}

@Article{Davey:2024xvd,
  author        = {Davey, Alex and Dias, Oscar J. C. and Gil, David Sola},
  journal       = {JHEP},
  title         = {{Strong Cosmic Censorship in Kerr-Newman-de Sitter}},
  year          = {2024},
  pages         = {113},
  volume        = {07},
  archiveprefix = {arXiv},
  doi           = {10.1007/JHEP07(2024)113},
  eprint        = {2404.03724},
  primaryclass  = {gr-qc},
}

@Article{Dias:2018etb,
  author        = {Dias, Oscar J. C. and Reall, Harvey S. and Santos, Jorge E.},
  journal       = {JHEP},
  title         = {{Strong cosmic censorship: taking the rough with the smooth}},
  year          = {2018},
  pages         = {001},
  volume        = {10},
  archiveprefix = {arXiv},
  doi           = {10.1007/JHEP10(2018)001},
  eprint        = {1808.02895},
  primaryclass  = {gr-qc},
}

@Article{Mishra:2020jlw,
  author        = {Mishra, Akash K. and Chakraborty, Sumanta},
  journal       = {Phys. Rev. D},
  title         = {{Strong cosmic censorship conjecture in higher curvature gravity}},
  year          = {2020},
  number        = {6},
  pages         = {064041},
  volume        = {101},
  archiveprefix = {arXiv},
  doi           = {10.1103/PhysRevD.101.064041},
  eprint        = {1911.09855},
  primaryclass  = {gr-qc},
}

@Article{Gan:2019jac,
  author        = {Gan, Qingyu and Guo, Guangzhou and Wang, Peng and Wu, Houwen},
  journal       = {Phys. Rev. D},
  title         = {{Strong cosmic censorship for a scalar field in a Born-Infeld{\textendash}de Sitter black hole}},
  year          = {2019},
  number        = {12},
  pages         = {124009},
  volume        = {100},
  archiveprefix = {arXiv},
  doi           = {10.1103/PhysRevD.100.124009},
  eprint        = {1907.04466},
  primaryclass  = {hep-th},
  reportnumber  = {CTP-SCU/2019012},
}

@Article{Cardoso:2018nvb,
  author        = {Cardoso, Vitor and Costa, Joao L. and Destounis, Kyriakos and Hintz, Peter and Jansen, Aron},
  journal       = {Phys. Rev. D},
  title         = {{Strong cosmic censorship in charged black-hole spacetimes: still subtle}},
  year          = {2018},
  number        = {10},
  pages         = {104007},
  volume        = {98},
  archiveprefix = {arXiv},
  doi           = {10.1103/PhysRevD.98.104007},
  eprint        = {1808.03631},
  primaryclass  = {gr-qc},
}

@Article{Li:2026zsg,
  author        = {Li, Peiyang and Wang, Mengjie and Jing, Jiliang},
  title         = {{Charged scalar and Dirac perturbations on a global monopole Reissner-Nordstr{\"o}m-de Sitter black hole: quasinormal modes and strong cosmic censorship}},
  year          = {2026},
  month         = {2},
  archiveprefix = {arXiv},
  eprint        = {2602.23083},
  primaryclass  = {gr-qc},
}

@Article{Ge:2018vjq,
  author        = {Ge, Boxuan and Jiang, Jie and Wang, Bin and Zhang, Hongbao and Zhong, Zhen},
  journal       = {JHEP},
  title         = {{Strong cosmic censorship for the massless Dirac field in the Reissner-Nordstrom-de Sitter spacetime}},
  year          = {2019},
  pages         = {123},
  volume        = {01},
  archiveprefix = {arXiv},
  doi           = {10.1007/JHEP01(2019)123},
  eprint        = {1810.12128},
  primaryclass  = {gr-qc},
}

@Article{Liu:2019rbq,
  author        = {Liu, Xiaoyi and Van Vooren, Stijn and Zhang, Hongbao and Zhong, Zhen},
  journal       = {JHEP},
  title         = {{Strong cosmic censorship for the Dirac field in the higher dimensional Reissner-Nordstrom--de Sitter black hole}},
  year          = {2019},
  pages         = {186},
  volume        = {10},
  archiveprefix = {arXiv},
  doi           = {10.1007/JHEP10(2019)186},
  eprint        = {1909.07904},
  primaryclass  = {hep-th},
}

@Article{Mo:2018nnu,
  author        = {Mo, Yuyu and Tian, Yu and Wang, Bin and Zhang, Hongbao and Zhong, Zhen},
  journal       = {Phys. Rev. D},
  title         = {{Strong cosmic censorship for the massless charged scalar field in the Reissner-Nordstrom{\textendash}de Sitter spacetime}},
  year          = {2018},
  number        = {12},
  pages         = {124025},
  volume        = {98},
  archiveprefix = {arXiv},
  doi           = {10.1103/PhysRevD.98.124025},
  eprint        = {1808.03635},
  primaryclass  = {gr-qc},
}

@Article{Guo:2019tjy,
  author        = {Guo, Hong and Liu, Hang and Kuang, Xiao-Mei and Wang, Bin},
  journal       = {Eur. Phys. J. C},
  title         = {{Strong Cosmic Censorship in Charged de Sitter spacetime with Scalar Field Non-minimally Coupled to Curvature}},
  year          = {2019},
  number        = {11},
  pages         = {891},
  volume        = {79},
  archiveprefix = {arXiv},
  doi           = {10.1140/epjc/s10052-019-7416-x},
  eprint        = {1905.09461},
  primaryclass  = {gr-qc},
}

@Article{Rahman:2018oso,
  author        = {Rahman, Mostafizur and Chakraborty, Sumanta and SenGupta, Soumitra and Sen, Anjan A.},
  journal       = {JHEP},
  title         = {{Fate of Strong Cosmic Censorship Conjecture in Presence of Higher Spacetime Dimensions}},
  year          = {2019},
  pages         = {178},
  volume        = {03},
  archiveprefix = {arXiv},
  doi           = {10.1007/JHEP03(2019)178},
  eprint        = {1811.08538},
  primaryclass  = {gr-qc},
}

@Article{Destounis:2018qnb,
  author        = {Destounis, Kyriakos},
  journal       = {Phys. Lett. B},
  title         = {{Charged Fermions and Strong Cosmic Censorship}},
  year          = {2019},
  pages         = {211--219},
  volume        = {795},
  archiveprefix = {arXiv},
  doi           = {10.1016/j.physletb.2019.06.015},
  eprint        = {1811.10629},
  primaryclass  = {gr-qc},
}

@Article{Hod:2018dpx,
  author        = {Hod, Shahar},
  journal       = {Nucl. Phys. B},
  title         = {{Strong cosmic censorship in charged black-hole spacetimes: As strong as ever}},
  year          = {2019},
  pages         = {636--645},
  volume        = {941},
  archiveprefix = {arXiv},
  doi           = {10.1016/j.nuclphysb.2019.03.003},
  eprint        = {1801.07261},
  primaryclass  = {gr-qc},
}

@Article{Konoplya:2022kld,
  author        = {Konoplya, R. A. and Zhidenko, A.},
  journal       = {JCAP},
  title         = {{How general is the strong cosmic censorship bound for quasinormal modes?}},
  year          = {2022},
  pages         = {028},
  volume        = {11},
  archiveprefix = {arXiv},
  doi           = {10.1088/1475-7516/2022/11/028},
  eprint        = {2210.04314},
  primaryclass  = {gr-qc},
}

@Article{Courty:2023rxk,
  author        = {Courty, Aubin and Destounis, Kyriakos and Pani, Paolo},
  journal       = {Phys. Rev. D},
  title         = {{Spectral instability of quasinormal modes and strong cosmic censorship}},
  year          = {2023},
  number        = {10},
  pages         = {104027},
  volume        = {108},
  archiveprefix = {arXiv},
  doi           = {10.1103/PhysRevD.108.104027},
  eprint        = {2307.11155},
  primaryclass  = {gr-qc},
}

@Article{Zhang:2023yco,
  author        = {Zhang, Ming and Jiang, Jie},
  journal       = {Sci. China Phys. Mech. Astron.},
  title         = {{Strong Cosmic Censorship in accelerating spacetime}},
  year          = {2023},
  number        = {8},
  pages         = {280412},
  volume        = {66},
  archiveprefix = {arXiv},
  doi           = {10.1007/s11433-023-2117-7},
  eprint        = {2302.04738},
  primaryclass  = {gr-qc},
}

@Article{Dias:2019ery,
  author        = {Dias, Oscar J. C. and Reall, Harvey S. and Santos, Jorge E.},
  journal       = {JHEP},
  title         = {{The BTZ black hole violates strong cosmic censorship}},
  year          = {2019},
  pages         = {097},
  volume        = {12},
  archiveprefix = {arXiv},
  doi           = {10.1007/JHEP12(2019)097},
  eprint        = {1906.08265},
  primaryclass  = {hep-th},
}

@Article{Singha:2022bvr,
  author        = {Singha, Chiranjeeb and Chakraborty, Sumanta and Dadhich, Naresh},
  journal       = {JHEP},
  title         = {{Strong cosmic censorship conjecture for a charged BTZ black hole}},
  year          = {2022},
  pages         = {028},
  volume        = {06},
  archiveprefix = {arXiv},
  doi           = {10.1007/JHEP06(2022)028},
  eprint        = {2203.07708},
  primaryclass  = {gr-qc},
}

@Article{Emparan:2020rnp,
  author        = {Emparan, Roberto and Toma{\v{s}}evi{\'c}, Marija},
  journal       = {JHEP},
  title         = {{Strong cosmic censorship in the BTZ black hole}},
  year          = {2020},
  pages         = {038},
  volume        = {06},
  archiveprefix = {arXiv},
  doi           = {10.1007/JHEP06(2020)038},
  eprint        = {2002.02083},
  primaryclass  = {hep-th},
}

@Article{Chrysostomou:2025qud,
  author        = {Chrysostomou, Anna and Cornell, Alan S. and Deandrea, Aldo and Park, Seong Chan},
  journal       = {Class. Quant. Grav.},
  title         = {{A note on strong cosmic censorship and its violation in Reissner{\textendash}Nordstr{\"o}m de Sitter black hole space-times}},
  year          = {2025},
  number        = {10},
  pages         = {107001},
  volume        = {42},
  archiveprefix = {arXiv},
  doi           = {10.1088/1361-6382/add5b4},
  eprint        = {2501.12968},
  primaryclass  = {gr-qc},
}

@Article{Tu:2025zeb,
  author        = {Tu, Zhiqin and Tang, Meirong and Xu, Zhaoyi},
  title         = {{Yang-Mills field modified RN black hole and the Strong Cosmic Censorship Conjecture}},
  year          = {2025},
  month         = {1},
  archiveprefix = {arXiv},
  eprint        = {2501.06409},
  primaryclass  = {gr-qc},
}

@Article{Ashtekar:2005qt,
  author        = {Ashtekar, Abhay and Bojowald, Martin},
  journal       = {Class. Quant. Grav.},
  title         = {{Quantum geometry and the Schwarzschild singularity}},
  year          = {2006},
  pages         = {391--411},
  volume        = {23},
  archiveprefix = {arXiv},
  doi           = {10.1088/0264-9381/23/2/008},
  eprint        = {gr-qc/0509075},
  reportnumber  = {IGPG-05-09-01, AEI-2005-132},
}

@Article{Bojowald:2018xxu,
  author        = {Bojowald, Martin and Brahma, Suddhasattwa and Yeom, Dong-han},
  journal       = {Phys. Rev. D},
  title         = {{Effective line elements and black-hole models in canonical loop quantum gravity}},
  year          = {2018},
  number        = {4},
  pages         = {046015},
  volume        = {98},
  archiveprefix = {arXiv},
  doi           = {10.1103/PhysRevD.98.046015},
  eprint        = {1803.01119},
  primaryclass  = {gr-qc},
}

@Article{Ashtekar:2018lag,
  author        = {Ashtekar, Abhay and Olmedo, Javier and Singh, Parampreet},
  journal       = {Phys. Rev. Lett.},
  title         = {{Quantum Transfiguration of Kruskal Black Holes}},
  year          = {2018},
  number        = {24},
  pages         = {241301},
  volume        = {121},
  archiveprefix = {arXiv},
  doi           = {10.1103/PhysRevLett.121.241301},
  eprint        = {1806.00648},
  primaryclass  = {gr-qc},
}

@Article{Yang:2022yvq,
  author        = {Yang, Si-Jiang and Zhang, Yu-Peng and Wei, Shao-Wen and Liu, Yu-Xiao},
  journal       = {JHEP},
  title         = {{Destroying the event horizon of a nonsingular rotating quantum-corrected black hole}},
  year          = {2022},
  pages         = {066},
  volume        = {04},
  archiveprefix = {arXiv},
  doi           = {10.1007/JHEP04(2022)066},
  eprint        = {2201.03381},
  primaryclass  = {gr-qc},
}

@Article{Fernando:2015fha,
  author        = {Fernando, Sharmanthie},
  journal       = {Int. J. Mod. Phys. D},
  title         = {{Regular black holes in de Sitter universe: scalar field perturbations and quasinormal modes}},
  year          = {2015},
  number        = {14},
  pages         = {1550104},
  volume        = {24},
  archiveprefix = {arXiv},
  doi           = {10.1142/S0218271815501047},
  eprint        = {1508.03581},
  primaryclass  = {gr-qc},
}

@Article{Frolov:2016pav,
  author        = {Frolov, Valeri P.},
  journal       = {Phys. Rev. D},
  title         = {{Notes on nonsingular models of black holes}},
  year          = {2016},
  number        = {10},
  pages         = {104056},
  volume        = {94},
  archiveprefix = {arXiv},
  doi           = {10.1103/PhysRevD.94.104056},
  eprint        = {1609.01758},
  primaryclass  = {gr-qc},
}

@Article{Carballo-Rubio:2019fnb,
  author        = {Carballo-Rubio, Ra{\'u}l and Di Filippo, Francesco and Liberati, Stefano and Visser, Matt},
  journal       = {Phys. Rev. D},
  title         = {{Geodesically complete black holes}},
  year          = {2020},
  pages         = {084047},
  volume        = {101},
  archiveprefix = {arXiv},
  doi           = {10.1103/PhysRevD.101.084047},
  eprint        = {1911.11200},
  primaryclass  = {gr-qc},
}

@Article{Poisson:1990eh,
  author  = {Poisson, Eric and Israel, W.},
  journal = {Phys. Rev. D},
  title   = {{Internal structure of black holes}},
  year    = {1990},
  pages   = {1796--1809},
  volume  = {41},
  doi     = {10.1103/PhysRevD.41.1796},
}

@Article{Carballo-Rubio:2018pmi,
  author        = {Carballo-Rubio, Ra{\'u}l and Di Filippo, Francesco and Liberati, Stefano and Pacilio, Costantino and Visser, Matt},
  journal       = {JHEP},
  title         = {{On the viability of regular black holes}},
  year          = {2018},
  pages         = {023},
  volume        = {07},
  archiveprefix = {arXiv},
  doi           = {10.1007/JHEP07(2018)023},
  eprint        = {1805.02675},
  primaryclass  = {gr-qc},
}

@Article{Carballo-Rubio:2021bpr,
  author        = {Carballo-Rubio, Ra{\'u}l and Di Filippo, Francesco and Liberati, Stefano and Pacilio, Costantino and Visser, Matt},
  journal       = {JHEP},
  title         = {{Inner horizon instability and the unstable cores of regular black holes}},
  year          = {2021},
  pages         = {132},
  volume        = {05},
  archiveprefix = {arXiv},
  doi           = {10.1007/JHEP05(2021)132},
  eprint        = {2101.05006},
  primaryclass  = {gr-qc},
  reportnumber  = {YITP-21-02},
}

@Article{Bonanno:2020fgp,
  author        = {Bonanno, Alfio and Khosravi, Amir-Pouyan and Saueressig, Frank},
  journal       = {Phys. Rev. D},
  title         = {{Regular black holes with stable cores}},
  year          = {2021},
  number        = {12},
  pages         = {124027},
  volume        = {103},
  archiveprefix = {arXiv},
  doi           = {10.1103/PhysRevD.103.124027},
  eprint        = {2010.04226},
  primaryclass  = {gr-qc},
}

@Article{Bonanno:2022jjp,
  author        = {Bonanno, Alfio and Khosravi, Amir-Pouyan and Saueressig, Frank},
  journal       = {Phys. Rev. D},
  title         = {{Regular evaporating black holes with stable cores}},
  year          = {2023},
  number        = {2},
  pages         = {024005},
  volume        = {107},
  archiveprefix = {arXiv},
  doi           = {10.1103/PhysRevD.107.024005},
  eprint        = {2209.10612},
  primaryclass  = {gr-qc},
}

@Article{Carballo-Rubio:2022kad,
  author        = {Carballo-Rubio, Ra{\'u}l and Di Filippo, Francesco and Liberati, Stefano and Pacilio, Costantino and Visser, Matt},
  journal       = {JHEP},
  title         = {{Regular black holes without mass inflation instability}},
  year          = {2022},
  pages         = {118},
  volume        = {09},
  archiveprefix = {arXiv},
  doi           = {10.1007/JHEP09(2022)118},
  eprint        = {2205.13556},
  primaryclass  = {gr-qc},
  reportnumber  = {YITP-22-53},
}

@Article{Franzin:2022wai,
  author        = {Franzin, Edgardo and Liberati, Stefano and Mazza, Jacopo and Vellucci, Vania},
  journal       = {Phys. Rev. D},
  title         = {{Stable rotating regular black holes}},
  year          = {2022},
  number        = {10},
  pages         = {104060},
  volume        = {106},
  archiveprefix = {arXiv},
  doi           = {10.1103/PhysRevD.106.104060},
  eprint        = {2207.08864},
  primaryclass  = {gr-qc},
}

@Article{Liu:2026ltw,
  author        = {Liu, Hongguang and Soranidis, Ioannis},
  title         = {{Probing mass inflation in polymerized vacuum regular black holes via colliding null shells}},
  year          = {2026},
  month         = {4},
  archiveprefix = {arXiv},
  eprint        = {2604.27897},
  primaryclass  = {gr-qc},
}

@Article{Cao:2023aco,
  author        = {Cao, Li-Ming and Li, Long-Yue and Wu, Liang-Bi and Zhou, Yu-Sen},
  journal       = {Eur. Phys. J. C},
  title         = {{The instability of the inner horizon of the quantum-corrected black hole}},
  year          = {2024},
  number        = {5},
  pages         = {507},
  volume        = {84},
  archiveprefix = {arXiv},
  doi           = {10.1140/epjc/s10052-024-12832-4},
  eprint        = {2308.10746},
  primaryclass  = {gr-qc},
  reportnumber  = {ICTS-USTC/PCFT-23-23},
}

@Article{Cao:2023par,
  author        = {Cao, Li-Ming and Li, Long-Yue and Liu, Xia-Yuan and Zhou, Yu-Sen},
  journal       = {Phys. Rev. D},
  title         = {{Appearance of the regular black hole with a stable inner horizon}},
  year          = {2024},
  number        = {6},
  pages         = {064083},
  volume        = {109},
  archiveprefix = {arXiv},
  doi           = {10.1103/PhysRevD.109.064083},
  eprint        = {2312.04301},
  primaryclass  = {gr-qc},
  reportnumber  = {ICTS-USTC/PCFT-23-39},
}

@Article{Li:2013sea,
  author        = {Li, Zilong and Bambi, Cosimo},
  journal       = {Phys. Rev. D},
  title         = {{Destroying the event horizon of regular black holes}},
  year          = {2013},
  number        = {12},
  pages         = {124022},
  volume        = {87},
  archiveprefix = {arXiv},
  doi           = {10.1103/PhysRevD.87.124022},
  eprint        = {1304.6592},
  primaryclass  = {gr-qc},
}

@Article{Fauzi:2025ldu,
  author        = {Fauzi, M. F. and Ramadhan, H. S. and Sulaksono, A. and Hasanuddin, H.},
  journal       = {Class. Quant. Grav.},
  title         = {{Imaging the destruction of a rotating regular black hole}},
  year          = {2025},
  number        = {22},
  pages         = {225012},
  volume        = {42},
  archiveprefix = {arXiv},
  doi           = {10.1088/1361-6382/ae1ac3},
  eprint        = {2503.07011},
  primaryclass  = {gr-qc},
}

@Article{Lin:2024beb,
  author        = {Lin, Jianhui and Zhang, Xiangdong and Bravo-Gaete, Mois{\'e}s},
  journal       = {Phys. Rev. D},
  title         = {{Mass inflation and strong cosmic censorship conjecture in the covariant quantum black hole}},
  year          = {2025},
  number        = {10},
  pages         = {106025},
  volume        = {111},
  archiveprefix = {arXiv},
  doi           = {10.1103/n7jv-crs9},
  eprint        = {2412.01448},
  primaryclass  = {gr-qc},
}

@Article{Hintz:2016jak,
  author        = {Hintz, Peter},
  title         = {{Non-linear stability of the Kerr-Newman-de Sitter family of charged black holes}},
  year          = {2016},
  month         = {12},
  archiveprefix = {arXiv},
  doi           = {10.1007/s40818-018-0047-y},
  eprint        = {1612.04489},
  primaryclass  = {math.AP},
}

@Article{Hintz:2016gwb,
  author        = {Hintz, Peter and Vasy, Andr{\'a}s},
  journal       = {Acta Math.},
  title         = {{The global non-linear stability of the Kerr-de Sitter family of black holes.}},
  year          = {2018},
  number        = {1},
  pages         = {1--206},
  volume        = {220},
  archiveprefix = {arXiv},
  doi           = {10.4310/ACTA.2018.v220.n1.a1},
  eprint        = {1606.04014},
  primaryclass  = {math.DG},
}

@Article{Costa:2017tjc,
  author        = {Costa, Jo{\~a}o L. and Gir{\~a}o, Pedro M. and Nat{\'a}rio, Jos{\'e} and Silva, Jorge Drumond},
  journal       = {Commun. Math. Phys.},
  title         = {{On the Occurrence of Mass Inflation for the Einstein{\textendash}Maxwell-Scalar Field System with a Cosmological Constant and an Exponential Price Law}},
  year          = {2018},
  number        = {1},
  pages         = {289--341},
  volume        = {361},
  archiveprefix = {arXiv},
  doi           = {10.1007/s00220-018-3122-z},
  eprint        = {1707.08975},
  primaryclass  = {gr-qc},
}

@Article{Costa:2014aia,
  author        = {Costa, Jo{\~a}o L. and Gir{\~a}o, Pedro M. and Nat{\'a}rio, Jos{\'e} and Silva, Jorge Drumond},
  title         = {{On the global uniqueness for the Einstein-Maxwell-scalar field system with a cosmological constant. Part 3: Mass inflation and extendibility of the solutions}},
  year          = {2014},
  month         = {6},
  archiveprefix = {arXiv},
  doi           = {10.1007/s40818-017-0028-6},
  eprint        = {1406.7261},
  primaryclass  = {gr-qc},
}

\end{document}